\documentclass{aa}  

\usepackage{graphicx}
\usepackage{txfonts}
\usepackage{lipsum}
\usepackage{subcaption}         % necessary for continued figures, example in section 3
\usepackage{lscape}             % to rotate a single page table, example in appendix.
\usepackage{placeins}           % useful with \FloatBarrier, to keep 
\usepackage[colorlinks=true, linkcolor=blue, citecolor=blue]{hyperref} % Borrar la final
\begin{document}

\title{Validating and improving two-fluid simulations of the magnetic field evolution in neutron star cores}
\subtitle{}

\author{F. Castillo\inst{1}\fnmsep\thanks{E-mail: francisco.castillo@umce.cl}
\and N. A. Moraga\inst{2}
\and M.\ E.\ Gusakov\inst{3}
\and J. A. Valdivia\inst{2}
\and A. Reisenegger\inst{1}
}

\institute{
Departamento de F\'{\i}sica, Facultad de Ciencias B\'asicas, Universidad Metropolitana de Ciencias de la Educaci\'on, Av. Jos\'e Pedro Alessandri 774, \~Nu\~noa, Santiago, Chile\\
\and Departamento de F\'{\i}sica, Facultad de Ciencias,
Universidad de Chile, Las Palmeras 3425, \~Nu\~noa, Santiago, Chile\\
\and Ioffe Institute, Politekhnicheskaya 26, St. Petersburg, 194021, Russia
}

\date{Received September 30, 20XX}

\abstract
%Optional, leave empty if necessary.  The heading “Context” is used when needed to give background information on the research conducted in the paper
{
This paper addresses the evolution of an axially symmetric magnetic field in the core of a neutron star. The matter in the core is modeled as a system of two fluids, namely neutrons and charged particles, with slightly different velocity fields, controlled by their mutual collisional friction. This problem was addressed in our previous work through the so-called ``fictitious friction'' approach.
}
% Mandatory. The objectives of the paper are defined here.
{
We study the validity of our previous work and improve it by comparing the fictitious friction approach to alternatives, making approximations that allow it to be applied to arbitrary magnetic field strengths and using realistic equations of state. 
} 
% Mandatory. The methods of the investigation are outlined here
{
We assume the neutron star crust to be perfectly resistive, so its magnetic field reacts instantaneously to changes in the core, in which we neglect the effects of Cooper pairing. We explore different approaches to solve the equations to obtain the velocities and chemical potential perturbations induced by a given, fixed magnetic field configuration in the core. 
We also present a new version of our code to perform time-evolving simulations and discuss the results obtained with it.
}
% Mandatory. The results are summarized here.
{
Our calculations without fictitious friction further confirm that bulk velocity is generally much greater than ambipolar velocity, leading to faster evolution. These findings align with those with fictitious friction, validating this approach.
We also find that, in the long term, the star evolves towards a barotropic ``Grad-Shafranov equilibrium,'' where the magnetic force is fully balanced by charged particle fluid forces. Qualitatively, the evolution and the final equilibrium are independent of the magnetic field strength $B$ and the equation of state considered.
The timescale to reach this equilibrium is proportional to $B^{-2}$ and becomes shorter for equations of state with a smaller gradient of the ratio between the densities of protons and neutrons.
}
%Optional, leave empty if necessary.  “Conclusions” can be used to explicit the general conclusions that can be drawn from the paper.
{}

\keywords{Stars: neutron -- Stars: magnetic field -- Magnetohydrodynamics (MHD)
 -- Methods: numerical}

\maketitle

\nolinenumbers

\section{Introduction}
\label{sec:intro}

Neutron stars (NSs) are categorized into different classes (e.~g., magnetars, rotation-powered pulsars, rotating radio transients, isolated thermal emitters, etc.) based on their observable properties \citep{Ascenzi2024_NeutronstarMeasurementsMultimessenger}.
Efforts towards a ``Grand Unification'' \citep{Kaspi2010_GrandUnificationNeutron} of their vast observational phenomena suggest that the magnetic field serves as a unifying factor \citep{Harding2013_NeutronStarZoo,Vigano2013_UnifyingObservationalDiversity}. 
Consequently, the evolution of their magnetic field continues to be a relevant and active area of research.
(For a recent review, see \citealt{Igoshev2021_EvolutionNeutronStar}.)

The current picture of magnetic field evolution in these objects is as follows:
Minutes after their formation, charged particles in their interior (mostly protons [$p$], electrons [$e$], and muons [$\mu$]), which feel the magnetic force, are strongly coupled to each other and to neutrons [$n$] by collisions, behaving essentially as a single fluid.
However, their joint motion, which is driven by the magnetic force, is strongly hindered by fluid forces (pressure and gravity forces) that arise from stable stratification
\citep{Pethick1992_TopicsPhysicsHigh,Reisenegger1992_NewClassGmodes,Goldreich1992_MagneticFieldDecay}. 

During this stage, sound and Alfvén waves propagate and are eventually damped, allowing the star to (very quickly) settle into a state of hydromagnetic quasi-equilibrium.
The nature of such equilibria has been extensively studied in the literature, 
yielding geometries that vary from ordered axisymmetric ``twisted-torus'' configurations with poloidal and toroidal components which stabilize each other \citep{Braithwaite2004_FossilOriginMagnetic,Braithwaite2006_EvolutionMagneticField, Braithwaite2006_StableMagneticFields} to disordered non-axisymmetric equilibrium magnetic fields \citep{Braithwaite2008_NonaxisymmetricMagneticEquilibria,Becerra2022_EvolutionRandomInitial}, depending on the initial radial distribution of magnetic energy in the star \citep{Braithwaite2008_NonaxisymmetricMagneticEquilibria} and the main energy dissipation mechanism invoked, namely viscosity versus magnetic diffusivity \citep{Becerra2022_EvolutionRandomInitial}. 

While the star cools, a crust is formed; thus, the magnetic evolution continues in two different regions: a solid crust and a fluid core, with the transition between them occurring at approximately half of the nuclear saturation density.
As discussed by \citet{Goldreich1992_MagneticFieldDecay}, the two long-term mechanisms that promote the evolution of the field in the crust are
Ohmic diffusion (a dissipative effect due to electron-ion collisions) and Hall drift (non-dissipative advection of the magnetic field by the electrical current).
The evolution of the magnetic field in this region, its effect on thermal evolution, and its potential impact on observations have been extensively studied in the literature (see, for instance, \citealt{Hollerbach2002_InfluenceHallDrift,Pons2007_MagneticFieldDissipation,Pons2009_MagnetothermalEvolutionNeutron,Vigano2012_NewCodeHalldriven,Gourgouliatos2013_HallEffectNeutron,Vigano2013_UnifyingObservationalDiversity,Gourgouliatos2014_HallAttractorAxially,Marchant2014_STABILITYHALLEQUILIBRIA,Lander2019_MagneticfieldEvolutionPlastically, Vigano2021_MagnetothermalEvolutionNeutron}), where recent efforts even involve full three-dimensional simulations \citep{Wood2015_ThreeDimensionalSimulation,Dehman2023_3DCodeMAgnetoThermal,Ascenzi2024_3DCodeMAgnetoThermal}.

The core, on the other hand, follows a different picture. 
Young NSs have high temperatures, which, in principle, allow the core matter to overcome stable stratification, while still moving as a single fluid, strongly coupled by collisions. At such temperatures, the chemical imbalances due to the magnetic force are continuously reduced by weak interaction processes (Urca reactions).
This erodes the initial quasi-equilibrium state, allowing the magnetic field and particles to move through a continuous sequence of quasi-equilibrium states as the chemical imbalances are reduced \citep{Reisenegger2005_MagneticFieldsNeutron,Reisenegger2009_StableMagneticEquilibria,Ofengeim2018_FastMagneticField}.
However, this so-called ``strong coupling regime'' is short-lived. The star passively cools down due to Urca reactions faster than it is reheated from the feedback due to the magnetic evolution (even for magnetar-strength fields), soon rendering Urca reactions inefficient at restoring chemical equilibrium. Moreover, the timescale of magnetic field evolution increases very quickly as temperature decreases ($t_B\propto T^{-6}$); thus, the magnetic field is found to evolve very little in this stage \citep{Moraga2024_MagnetothermalEvolutionCores}.
As noted by \citet{Moraga2024_MagnetothermalEvolutionCores}, the latter implies that the kind of barotropic equilibrium configurations often used in the literature as initial configurations for proto-NSs (e.~g., \citealt{Thompson1996_SoftGammaRepeaters, Lander2021_MagneticFieldsLatestage}), 
which are solutions of the ``Grad-Shafranov'' (GS) equation (see Sect.~\ref{sec:evolution:GS} for details), cannot be reached in this early stage of evolution.
However, as shown by \citet{Castillo2017_MagneticFieldEvolution,Castillo2020_TwofluidSimulationsMagnetic}, GS equilibrium configurations are expected to appear at later times in the so-called ``weak coupling regime,'' which we discuss next.

As the star cools further, the collisional coupling between charged particles and neutrons is reduced, allowing for a motion of charged particles relative to neutrons in a dissipative process called ambipolar diffusion \citep{Pethick1992_TopicsPhysicsHigh, Goldreich1992_MagneticFieldDecay}.
The velocity of this relative motion is proportional to the imbalance between the magnetic force and the charged particle fluid forces it induces, and also inversely proportional to the collision rate between charged particles and neutrons.
Such motion erodes the initial hydromagnetic quasi-equilibrium, promoting the evolution of the magnetic field and fluid forces as they continually re-adjust, moving through a continuum of successive quasi-equilibria.
In \citet{Castillo2020_TwofluidSimulationsMagnetic}, we reported two-fluid numerical simulations of ambipolar diffusion considering a core of normal (non-Cooper-paired) $npe$ matter in axial symmetry.
We showed that, as ambipolar diffusion acts, the fluid forces due to neutrons and charged particles adjust, with the former going to zero and the latter eventually fully balancing the magnetic force.
Thus, an equilibrium is reached, which can only be very slowly eroded by the combination of Hall drift and Ohmic dissipation.
Since charged particles are fully decoupled from neutrons, this corresponds to a single-fluid barotropic equilibrium configuration, which is a solution of the GS equation.

Ambipolar diffusion is arguably the main effect promoting the magnetic field evolution in the core (in that region, it is certainly more effective than Hall drift and Ohmic decay as conductivity is much higher). Consequently, it has been proposed as a plausible alternative mechanism explaining the weak magnetic fields of millisecond pulsars \citep{Cruces2019_WeakMagneticField}. It is also sometimes invoked to explain the activity of magnetars as it strongly depends on the magnetic field strength \citep{Thompson1995_SoftGammaRepeaters,Thompson1996_SoftGammaRepeaters,Mereghetti2008_StrongestCosmicMagnets,Beloborodov2016_MagnetarHeating,Tsuruta2023_AmbipolarHeatingMagnetars}, although recent evidence suggests otherwise (Moraga et al., in preparation).
Thus, its impact on the long-term evolution of the field is a matter of interest.

The method to self-consistently solve the relevant equations for $npe$ matter (see Sect.~\ref{sec:model}) was introduced by \citet{Gusakov2017_EvolutionMagneticField} and later refined by \citet{Ofengeim2018_FastMagneticField}.
The latter authors presented a detailed scheme to semi-analytically solve the relevant system of equations for a given initial magnetic field configuration and demonstrated that, in general, the bulk motion of the particle fluids can be much larger than the relative velocity between species.
This finding was confirmed by the numerical simulations of the magnetic field evolution presented in \citet{Castillo2020_TwofluidSimulationsMagnetic}.
Taking this into account makes the evolution significantly faster than previously estimated (e.~g., \citealt{Goldreich1992_MagneticFieldDecay, Thompson1995_SoftGammaRepeaters,Beloborodov2016_MagnetarHeating,Castillo2017_MagneticFieldEvolution,Passamonti2017_RelevanceAmbipolarDiffusion}).

The semi-analytical scheme proposed by \citet{Gusakov2017_EvolutionMagneticField} and \citet{Ofengeim2018_FastMagneticField}, while useful to calculate the velocities and chemical potential perturbations induced by an analytically given magnetic field, has limited applicability in long-term numerical simulations, as the latter involve iterating over magnetic field data stored in tables, not given as explicit analytical expressions.
The authors mention that using magnetic field configurations in the form of tables in their scheme produces unreliable results; this limitation is attributed to the high number of spatial derivatives involved in the calculation (see Sect.~\ref{sec:model}), which yield numerical errors that accumulate if iterated over time (private communication).

Alternatively, in \citet{Castillo2020_TwofluidSimulationsMagnetic}, we followed a different approach. 
Instead of directly solving the relevant equations, we introduced a fictitious friction (FF) force acting on the neutrons.
The FF approach can be regarded as a numerical trick that greatly simplifies the numerical calculation of the relevant velocities and yields a very good approximation to the solution of the relevant equations (see a detailed discussion in Sect.~\ref{sec:fictitious_friction}).
This approach allowed us to report the first two-fluid, axially symmetric, numerical simulations of ambipolar diffusion, 
correctly capturing the bulk motion of neutrons and charged particles,
which cannot be done with the one-fluid approximation usually found in the literature
(e.~g., \citealt{Goldreich1992_MagneticFieldDecay,Castillo2017_MagneticFieldEvolution,Igoshev2023_ThreedimensionalMagnetothermalEvolution}).
While definitely a step forward, that work had some features that may have been not entirely convincing to some readers:
\begin{enumerate}[I.]
\item The approach used involved two timescales that are very different for realistic magnetic fields. To capture both in the simulations, we had to consider unrealistically high values of the magnetic field ($\gtrsim 10^{17}\text{G}$).

\item The FF approach is conceptually not trivial, and, at first glance, it is not evident that it yields the correct velocities fields.

\item The toy model equation of state (EoS) used (see Sect.~\ref{sec:EoS}) correctly captured the qualitative impact of stable stratification on the long-term magnetic evolution. However, the question remained if considering more realistic models would significantly affect the results reported.
\end{enumerate}

In this paper, we compare different methods to solve the relevant equations, including the FF approach and semi-analytical approaches based on the explicit scheme of \citet{Gusakov2017_EvolutionMagneticField} and \citet{Ofengeim2018_FastMagneticField}, as well as a pseudo-spectral method. 
Moreover, we present results of an updated version of the code used in \citet{Castillo2020_TwofluidSimulationsMagnetic} that addresses numerals I. and III. and present arguments to convince the reader about the correctness and applicability of the FF approach (thus addressing numeral II.).

This paper is organized as follows. 
In Sect.~\ref{sec:model}, we discuss the physical model and construct the equations to be solved in axial symmetry. 
In Sect.~\ref{sec:semi-anallytical} and Sect.~\ref{sec:fictitious_friction}, we describe different approaches to obtain the relevant velocities and chemical potential perturbations.
In Sect.~\ref{sec:EoS} and Sect.~\ref{sec:magnetic_models}, we present the different stellar and magnetic field models used in this work.
In Sect.~\ref{sec:evolution}, we describe the relevant equations for the magnetic field evolution, discuss the relevant time scales, and outline the mechanisms that promote the dissipation of the magnetic field.
Our results are presented in Sect.~\ref{sec:results}, where we compare the different approaches to obtain the neutron velocity and use the FF approach to perform long-term time-evolving simulations. We check if the simulations are in agreement with the expected timescales and final equilibrium configurations.
Finally, in Sect.~\ref{sec:conclusions}, our results are summarized and our conclusions are outlined.

\section{Physical model}
\label{sec:model}

Following the discussion of \citet{Castillo2020_TwofluidSimulationsMagnetic}, we model the NS core as a fluid composed of normal (not superfluid or superconducting) neutrons, protons, and electrons, in axial symmetry.
In the absence of a magnetic field, these are in a spherically symmetric hydrostatic and chemical equilibrium, having number densities $n_i(r)$ and chemical potentials $\mu_i(r)$ ($i=n,p,e$), respectively.
Note that charge neutrality implies $n_c\equiv n_p=n_e$, and chemical equilibrium implies $\mu\equiv\mu_n=\mu_p+\mu_e$.

For strong enough magnetic fields, the relative velocity of protons and electrons (related to the electric current density) is much smaller than their bulk velocity \citep{Ofengeim2018_FastMagneticField}, so all charged particles are assumed to move together with velocity $\vec{v}_c$, carrying along the magnetic flux according to Faraday's law,
\begin{equation}
    \frac{\partial\vec{B}}{\partial t}=\vec{\nabla}\times(\vec{v}_c\times\vec{B}), \label{eq:faraday}
\end{equation}
whereas the neutrons are allowed to move with a different velocity $\vec{v}_n$. The main challenge in simulating this process is to obtain the velocity fields $\vec{v}_c$ and $\vec{v}_n$ at each time step.

In the presence of a magnetic field $\vec{B}$, the magnetic force causes small perturbations to the number densities of neutrons, $\delta n_n$, and charged particles, $\delta n_c\equiv\delta n_p=\delta n_e$, which evolve following the continuity equations
\begin{gather}
\frac{\partial\delta n_n}{\partial t} + \vec{\nabla}\cdot\left(n_n\vec{v}_n\right) = \lambda\Delta\mu\label{eq:continuity_n}\,,\\
\frac{\partial\delta n_c}{\partial t} + \vec{\nabla}\cdot\left(n_c\vec{v}_c\right) = -\lambda\Delta\mu \,. \label{eq:continuity_c}
\end{gather} 
Here, $\lambda$ is a temperature-dependent parameter; $\Delta\mu\equiv\delta\mu_c-\delta\mu_n$; $\delta\mu_c\equiv\delta\mu_p+\delta\mu_e$; and $\delta\mu_n$, $\delta\mu_p$, and $\delta\mu_e$ are the chemical potential perturbations of the neutrons, protons, and electrons, respectively.
The terms on the right-hand side of the equations represent the net rates per unit volume at which weak interactions convert protons and electrons into neutrons, and vice versa; assuming $|\Delta\mu| \ll k_BT$, where $k_B$ is the Boltzmann constant and $T$ is the absolute temperature \citep{Goldreich1992_MagneticFieldDecay}.
In what follows, we focus on the low-temperature, ``weak-coupling'' regime, where $\lambda\approx 0$. 
Chemical potential perturbations are related to the number density perturbations $\delta n_n$ and $\delta n_c$ by
\begin{gather}
\delta\mu_n = K_{nn}\delta n_n + K_{nc}\delta n_c \label{eq:dmun}\,,\\
\delta\mu_c = K_{cn}\delta n_n + K_{cc}\delta n_c \label{eq:dmuc}\,,
\end{gather}
where $K_{ij}\equiv\partial\mu_i/\partial n_j$ ($i,j=c,n$). The off-diagonal element $K\equiv K_{nc}=K_{cn}$ accounts for strong interactions between protons and neutrons.

For any realistic magnetic field, the evolution timescale of such perturbations will be many orders of magnitude shorter than the magnetic evolution timescale.
Therefore, in time-evolving simulations, resolving the time-evolution timescale of the density perturbations and magnetic field for realistic magnetic field strength is computationally prohibitive.
However, as shown by \citet{Gusakov2017_EvolutionMagneticField}, the time derivatives in the continuity Eqs.~\eqref{eq:continuity_n}, and  \eqref{eq:continuity_c} can be safely neglected in the study of the long-term evolution, as the density perturbation of charged particles and neutrons will almost instantaneously adjust to the current configuration of the magnetic field. 
Thus, we reduce the continuity equations to
\begin{gather}
	\vec{\nabla}\cdot\left(n_n\vec{v}_n\right) = 0\label{eq:continuity_n2}\,,\\
	\vec{\nabla}\cdot\left(n_c\vec{v}_c\right) = 0 \,. \label{eq:continuity_c2}
\end{gather}

Each particle species feels fluid forces due to their weight and degeneracy pressure gradient. Namely, the fluid forces on the neutrons and charged particles are, respectively, $\vec{f}_{i} =-n_i\vec{\nabla}\delta\mu_i - n_i(\delta\mu_i/c^2)\vec{\nabla}\Phi$, for $i=n,c$; where $\Phi(r)$ is the gravitational potential.
Note that the unperturbed star is in hydrostatic equilibrium, satisfying
\begin{equation}
\vec{\nabla}\mu+\mu\vec{\nabla}\Phi/c^2=0 \,. \label{eq:hydrostatic equilibrium}
\end{equation}
Thus, by the introduction of the relative chemical potential perturbations $\chi_i=\delta\mu_i/\mu$ ($i=n,c$), fluid forces can be rewritten in a more compact form
\begin{gather}
\vec{f}_{n}=-n_n\mu\vec{\nabla}\chi_n\label{eq:neutron force} \,,\\
\vec{f}_{c}=-n_c\mu\vec{\nabla}\chi_c\label{eq:charged force} \,.
\end{gather}
Charged particles also feel the Lorentz force,
\begin{equation}
\vec{f}_{B}=\frac{(\vec{\nabla}\times\vec{B})\times\vec{B}}{4\pi} \label{eq:Lorentz force} \,.
\end{equation}

An underlying assumption is that the star very quickly reaches a state
of hydromagnetic quasi-equilibrium, in which all the forces acting on a fluid element are close to balance, yielding
\begin{equation}
\vec{f}_{B}-n_n\mu\vec{\nabla}\chi_n-n_c\mu\vec{\nabla}\chi_c=0\,.\label{eq:force_balance_1}
\end{equation}
Consequently, in the long term, the magnetic field evolves moving through a continuum of successive hydromagnetic quasi-equilibria.
In axial symmetry, the gradients $\vec{\nabla}\chi_i$ cannot have a toroidal component, therefore, Eq.~\eqref{eq:force_balance_1} imposes a strong constraint on the magnetic field, which must satisfy $f_{B,\phi}=0$.

The relative velocity of the charged particles with respect to the neutrons is the so-called ``ambipolar velocity,'' which is obtained from the force balance on the charged particles as
\begin{equation}
\vec{v}_{ad}\equiv\vec{v}_c-\vec{v}_n = \frac{\vec{f}_{B}-n_c\mu\vec{\nabla}\chi_c}{\gamma_{cn}n_c n_n}  \label{eq:v_ambipolar}\,.\\
\end{equation}
where $\gamma_{cn}$ parametrizes the collision rate between charged particles and neutrons.

Fluid forces are written in terms of gradients of the chemical potential perturbations (see Eqs.~\ref{eq:neutron force} and \ref{eq:charged force}).
Thus, there are additive constants in $\chi_{n}$ and $\chi_{c}$ that do not play a role in the evaluation of the velocities.
If needed, these can be fixed using integral conditions corresponding to the conservation of charged particles and neutrons, i.~e.,
\begin{gather}
\int_{\mathcal{V}_\text{core}}\delta n_n \,d\mathcal{V} =  \int_{\mathcal{V}_\text{core}} \mu \frac{K_{cc}\chi_n-K\chi_c}{K_{nn}K_{cc}-K^2} \,d\mathcal{V}  = 0\,,\label{eq:cons_n}\\
\int_{\mathcal{V}_\text{core}}\delta n_c \,d\mathcal{V} =  \int_{\mathcal{V}_\text{core}} \mu \frac{K_{nn}\chi_c-K\chi_n}{K_{nn}K_{cc}-K^2} \,d\mathcal{V}  = 0 \,,\label{eq:cons_c}
\end{gather}
where the density perturbations are obtained inverting Eqs.~\eqref{eq:dmun} and \eqref{eq:dmuc}, and $\mathcal{V}_\text{core}$ is the volume of the core.

As stated before, we restrict ourselves to axial symmetry, so the magnetic field can be decomposed as
\begin{equation}
\vec{B}=\vec{\nabla}\alpha\times\vec{\nabla}\phi + \beta\vec{\nabla}\phi \,. \label{eq:BAlfaBeta}
\end{equation}
The scalar potentials $\alpha(r,\theta)$ and $\beta(r,\theta)$ generate the poloidal and toroidal magnetic field, respectively, and $\vec{\nabla}\phi=\hat{\phi}/(r\sin\theta)$.

At each time step, the Lorentz force $\vec{f}_{B}$ is known, but the velocity fields $\vec{v}_{i}$ and the scalar fields $\chi_i$ for $i=n,c$ must be obtained from the self-consistent solution of Eqs.~\eqref{eq:continuity_n2}, \eqref{eq:continuity_c2}, \eqref{eq:force_balance_1}, and \eqref{eq:v_ambipolar}. Different approaches to this challenge are discussed in Sect.~\ref{sec:semi-anallytical}.
To perform numerical calculations using such approaches, the equations must be normalized appropriately.
The units in which the different variables are measured are given in Table~\ref{tab:normalization}.

\begin{table}
    \setlength{\tabcolsep}{15pt}
	\caption{Normalization of the relevant variables, where $\langle \vec{a}\rangle$ denotes the rms of a vector field $\vec{a}$ in the volume $\mathcal{V}_\text{core}$ of the core, $\langle \vec{a}\rangle=\sqrt{\frac{1}{\mathcal{V}_\text{core}}\int_{\mathcal{V}_\text{core}}|\vec{a}|^2\,d\mathcal{V}}$.
    The core radius $R$ and the different quantities evaluated at the center of the star ($n_0$, $\mu_0$, and $\gamma_0$) are dependent on the stellar model (see Sect.~\ref{sec:EoS}). Their values are given in Table~\ref{tab:stellar parameters}.
    }
	\label{tab:normalization}
	\centering
	\begin{tabular}{cc} % 
		\hline\hline
  		Variable      & Unit                    \\
		\hline
		$\vec{B}$ &  $B_0=\langle\vec{B}(t=0)\rangle$  \\ 
		$r$ &  $R$ (core radius)     \\ 
        $\vec\nabla$ &  $1/R$    \\ 
		$\alpha$ &  $B_0R^2$ (see Eq.~\ref{eq:BAlfaBeta}) \\ 
		$\beta$ &  $B_0R$ (see Eq.~\ref{eq:BAlfaBeta}) \\ 
		$\mu$ &  $\mu_0\equiv\mu(r=0)$    \\ 
		$n_n, n_c$ &  $n_{0}\equiv n_c(r=0)$   \\ 
        $\chi_n, \chi_c$ &  $\chi_0\equiv B_0^2/4\pi n_0\mu_0$  \\ 
		$K_{nn},K_{cc},K$ &  $\mu_0/n_0$    \\ 
		$\gamma_{cn}$ &  $\gamma_{0}\equiv\gamma_{cn}(r=0,T=\,10^8 \, \text{K})$    \\ 
        $\zeta$ &  $\gamma_{0}n_{0}$  \\ 
        $t$ &  $t_{0}=4\pi\gamma_{0}n_0^2R^2/B_0^2$  \\ 
        $\vec{v}_{c}, \vec{v}_{n}$ &  $R/t_{0}$  \\ 
		\hline
	\end{tabular}
\end{table}

\section{Semi-analytical approaches}
\label{sec:semi-anallytical}

\subsection{Explicit approach}
\label{sec:semi-anallytical:explicit}

The set of Eqs.~\eqref{eq:continuity_n2}, \eqref{eq:continuity_c2}, \eqref{eq:force_balance_1}, and \eqref{eq:v_ambipolar} can be solved for an analytically given magnetic field configuration in axial symmetry for which $f_{B,\phi}=0$. 
A semi-analytical procedure to solve the set of equations was described by \citet{Gusakov2017_EvolutionMagneticField} and more in depth by \citet{Ofengeim2018_FastMagneticField}. Their steps to obtain the poloidal component of the velocities and chemical potential perturbations are summarized here, adopting the notation and approximations used in the present paper.

The poloidal part of Eq.~\eqref{eq:force_balance_1} implies scalar equations for the radial and polar components, respectively.
For a given magnetic force $\vec{f}_{B}$, these equations fix $\chi_n$ and $\chi_c$ up to a radial function for each, which can be fixed by taking the appropriate boundary conditions. 
However, these conditions are not trivial and must be set for $\vec{v}_n$.
To proceed, we split the relative chemical imbalances into two components: $\chi_n(r,\theta)=X_n(r,\theta)+\chi_{n,0}(r)$, and $\chi_c(r,\theta)=X_c(r,\theta)+\chi_{c,0}(r)$, where $\chi_{n,0}=\hat{\mathrm{P}}_{0}\{\chi_{n}\}$ and $\chi_{c,0}=\hat{\mathrm{P}}_{0}\{\chi_{c}\}$. Here, we introduced the operator $\hat{\mathrm{P}}_{l}$ that extracts the coefficient of the $l$'th Legendre component of its argument,
\begin{equation}
    \hat{\mathrm{P}}_{l} \{f(r,\theta)\} \equiv \dfrac{2l+1}{2}\int^{\pi}_{0}\, f(r,\theta) \,P_{l}(\cos\theta) \sin \theta d\theta,
\end{equation}
where $f(r,\theta)$ is a given function of $r$ and $\theta$,
and $P_{l}(\cos\theta)$ is a Legendre polynomial of degree $l$. The relation between these radial functions can be obtained by taking the radial component of Eq.~\eqref{eq:force_balance_1} and projecting its $0$'th Legendre component
\begin{equation}
\hat{\mathrm{P}}_{0}\{f_{B,r}\} - n_c\mu\,\chi_{c,0}' - n_n\mu\,\chi_{n,0}' = 0 \,, \label{eq:Fbalance_r} 
\end{equation}
where prime ($'$) denotes derivative with respect to $r$. Replacing the decomposition of $\chi_c$ and $\chi_n$ into Eq.~\eqref{eq:force_balance_1}, together with the latter equation, yields
\begin{equation}
\vec{F}_{B}-n_n\vec{\nabla} X_n-n_c\vec{\nabla} X_c=0\,, \label{eq:force_balance_X}\\
\end{equation}
where $\vec{F}_{B}\equiv(\vec{f}_{B}-\hat{\mathrm{P}}_{0}\{f_{B,r}\}\hat r)/\mu$, and $\hat r$ is the radial unit vector. 
Note that, by defining $\epsilon\equiv n_n X_n+n_c X_c$ and $\delta \equiv n_n' X_n+n_c' X_c$, Eq.~(\ref{eq:force_balance_X}) can be rewritten as $\vec{\nabla}\epsilon-\delta\hat r=\vec{F}_{B}$.
The solution for $\epsilon$ and $\delta$ can be easily obtained by first integrating its polar component with respect to $\theta$. Thus,
\begin{gather}
\epsilon(r,\theta)=\int_{0}^\theta \,rF_{B,\theta}\,d\theta + \epsilon(r,0) \,, \label{eq:epsilon_sol}\\
\delta(r,\theta)=\frac{\partial\epsilon}{\partial r}-F_{B,r} \,, \label{eq:delta_sol}
\end{gather}
where the latter comes from the radial component and 
$\epsilon(r,0)= - \hat{\mathrm{P}}_{0}\{\int_{0}^\theta \,rF_{B,\theta}\,d\theta\}$
is chosen so $\hat{\mathrm{P}}_{0}\{\epsilon\}=0$.
In terms of these variables, we can reconstruct
\begin{gather}
X_n= \frac{n_c'\epsilon-n_c\delta}{n_n^2(n_c/n_n)'}  \,, \label{eq:Xn_ed}\\
X_c= -\frac{n_n'\epsilon-n_n\delta}{n_n^2(n_c/n_n)'} \,,
\label{eq:Xc_ed}
\end{gather}
which also satisfy $\hat{\mathrm{P}}_{0}\{X_n\}=\hat{\mathrm{P}}_{0}\{X_c\}=0$.

To fully determine $\chi_{n}$ and $\chi_{c}$, we must also evaluate $\chi_{n,0}$ and $\chi_{c,0}$.
Note that by integrating Eq.~\eqref{eq:continuity_n2} over the volume $\mathcal{V}(r)$ of any centered sphere of radius $r$, we obtain
\begin{equation}
    0=\int_{\mathcal{V}(r)} \vec{\nabla}\cdot\left(n_n\vec{v}_n\right)\,d\mathcal{V}=2\pi r^2 n_n\int_0^\pi  v_{n,r}(r,\theta) \sin\theta\,d\theta\,,
\end{equation}
implying $\hat{\mathrm{P}}_{0}\{v_{n,r}\}=0$. Analogously, from Eq.~\eqref{eq:continuity_c2} we obtain $\hat{\mathrm{P}}_{0}\{v_{c,r}\}=0$, and thus also $\hat{\mathrm{P}}_{0}\{v_{ad,r}\}=0$.
On the other hand, replacing Eq.~\eqref{eq:force_balance_1} and $\chi_n=X_n+\chi_{n,0}$ into Eq.~\eqref{eq:v_ambipolar}, yields 
\begin{equation}
n_c\vec{v}_{ad}=\frac{\mu}{\gamma_{cn}}\left(\vec{\nabla}X_n+\chi_{n,0}'\hat r\right)  \label{eq:v_ambipolar2}\,.\\
\end{equation}
Taking the radial component of the latter and projecting its $0$'th Legendre component implies 
\begin{equation}
\chi_{n,0}'=0\,. \label{eq:cond_chi_n0}
\end{equation}
This, together with Eq.~\eqref{eq:Fbalance_r}, implies
\begin{equation}
\chi_{c,0}'=\frac{\hat{\mathrm{P}}_{0}\{f_{B,r}\}}{n_c\mu}\,. \label{eq:cond_chi_c0}
\end{equation}
These define $\chi_{n,0}$ and $\chi_{c,0}$ from $f_{B,r}$ up to additive constants that do not affect the calculation of the velocities. If needed, they can be fixed using 
Eqs.~\eqref{eq:cons_n} and \eqref{eq:cons_c}.

Finally, the velocities can be obtained as follows:
$\vec{v}_{ad}$, which is purely poloidal, can be obtained from Eqs.~\eqref{eq:v_ambipolar2} and \eqref{eq:cond_chi_n0}
\begin{equation}
\vec{v}_{ad}=\frac{\mu \vec{\nabla}X_n}{n_c\gamma_{cn}}  \label{eq:v_ambipolar23}\,,\\
\end{equation}
whereas $\vec{v}_{n}$ follows by writing  $\vec{v}_c=\vec{v}_n+\vec{v}_{ad}$ and replacing Eq.~\eqref{eq:v_ambipolar23} into Eq.~\eqref{eq:continuity_c2}, which implies
\begin{equation}
\vec{\nabla}\cdot\left(n_c\vec{v}_n\right) + \vec{\nabla}\cdot\left(\frac{\mu}{\gamma_{cn}}\vec{\nabla} X_n \right) = 0 \,. \label{eq:continuity_c6}
\end{equation}
By multiplying Eq.~\eqref{eq:continuity_n2} by $n_c$, Eq.~\eqref{eq:continuity_c6} by $n_n$ and then subtracting them, we obtain, after some algebra
\begin{equation}
v_{n,r}=-\frac{1}{n_n(n_c/n_n)'}\vec{\nabla}\cdot\left(\frac{\mu}{\gamma_{cn}}\vec{\nabla} X_n \right) \,.\label{eq:vnr_anl}\\
\end{equation}
The polar component can be obtained from Eq.~\eqref{eq:continuity_n2}
\begin{equation}
v_{n,\theta}=-\frac{1}{r\sin\theta\, n_n}\frac{\partial}{\partial r}\left( r^2 n_n\int_0^\theta v_{n,r}\sin\theta\,d\theta\right) \,, \label{eq:vnth_anl}
\end{equation}
where the limits of the integral are chosen to be consistent with conditions arising from axial symmetry, i.~e., $v_{n,\theta}(r,0)=v_{n,\theta}(r,\pi)=0$.

On the other hand, the toroidal magnetic force must remain null ($\partial f_{B,\phi}/\partial t=0$), which imposes a quite complicated condition that can, in principle, be used to obtain $v_{n,\phi}$ (see \citealt{Ofengeim2018_FastMagneticField}). So far, no feasible numerical implementation to obtain $v_{n,\phi}$ for an arbitrary magnetic field configuration has been proposed.

Summarizing, the steps to obtain the poloidal components of $\vec{v}_{n}$ and $\vec{v}_{c}$ for an analytically given magnetic field satisfying $f_{B,\phi}=0$ are:
\begin{enumerate}
\item Evaluate $\epsilon$ and $\delta$ from the magnetic force using Eqs.~\eqref{eq:epsilon_sol} and \eqref{eq:delta_sol}.
\item Evaluate $X_n$ from Eq.~\eqref{eq:Xn_ed}.
\item Evaluate $\vec{v}_{ad}$ from Eq.~\eqref{eq:v_ambipolar23}.
\item Evaluate $v_{n,r}$ and $v_{n,\theta}$ from Eqs.~\eqref{eq:vnr_anl} and \eqref{eq:vnth_anl}, respectively.
\item Evaluate $\vec{v}_{c\text{, Pol}}=\vec{v}_{n\text{, Pol}}+\vec{v}_{ad\text{, Pol}}$,
\end{enumerate}
where the subscript `Pol' refers to the poloidal part of a vector field, e.g., $\vec{v}_{n\text{, Pol}}=v_{n,r}\hat r + v_{n,\theta}\hat \theta$.

We note that this procedure completely defines the velocity fields everywhere in the NS core, leaving no room for imposing boundary conditions at the crust-core interface. Thus, any such conditions, such as no flux of particles through the interface ($v_{n,r}(r=R)=v_{c,r}(r=R)=0$), require constraining the magnetic field \citep{Ofengeim2018_FastMagneticField}, and it is unclear whether these constraints will be maintained as the magnetic field evolves in time. 

In Sect.~\ref{sec:results:comparison}, we show plots of the velocities, which are computed using the procedure described above. All steps are evaluated using Wolfram Mathematica\footnote{www.wolfram.com} software, which produces involved symbolic expressions for the velocities and chemical potential perturbations in terms of the coordinates $r,\theta$, the magnetic force, the functions $n_c(r), n_n(r),\mu(r),\gamma_{cn}(r)$, and their radial derivatives.
The latter background functions are evaluated from numerical data computed from the EoS.

\subsection{Pseudo-spectral approach}
\label{sec:semi-anallytical:spectral}

In addition to the semi-analytical procedure outlined in Sect.~\ref{sec:semi-anallytical:explicit}, the system of Eqs.~\eqref{eq:continuity_n2}, \eqref{eq:continuity_c2}, \eqref{eq:force_balance_1}, and \eqref{eq:v_ambipolar} can be solved using a pseudo-spectral approach.
An optimal implementation of such an algorithm requires rewriting some of the previous equations in a spectral-friendly manner.

Spectral codes expand scalar and vector fields using polynomial bases for the radial and angular domains within a spherical region. The angular component is typically expanded in spherical harmonics as they satisfy the regularity conditions along the axis. In axial symmetry, these reduce to the Legendre polynomials $P_l(\cos\theta)$. Consequently, the chemical perturbations read as
\begin{equation}\label{eqChiEspectral}
    \chi_{i}(r,\theta) = \sum_{l = 0}^{n_{\theta}} \chi_{i,l}(r)P_{l}(\cos \theta), \quad (i=n,c)\,,
\end{equation}
where $n_{\theta}$ represents the truncation order of the polynomial basis in the angular domain, and $\chi_{i,l}(r)=\hat{\mathrm{P}}_{l}\{\chi_{i}\}$; thus $X_i=\sum_{l\geq 1}\chi_{i,l}(r)P_{l}(\cos \theta) $.
For simplicity, we have omitted the explicit expansion of $\chi_{i,l}(r)$ in radial polynomials.
The solution procedure is the following:
\begin{enumerate}
    \item First, we compute the chemical potential perturbations.
    Instead of using Eqs.~\eqref{eq:Xn_ed} and Eqs.~\eqref{eq:Xc_ed} to obtain the $\{\chi_{i,l}\}_{l\geq 1}$, we proceed in the following manner.
    To obtain $\chi_c$, we divide Eq.~(\ref{eq:force_balance_1}) by $\mu n_n$ and then take the curl
    \begin{equation}\label{eqCurlChic}
    \vec \nabla \times \left[ \left( \frac{n_c}{n_n} \right)'  \chi_c \hat{r}\right] = - \vec{\nabla}\times \left(\dfrac{ \vec{f}_{B}}{\mu n_n}\right)\,.
    \end{equation}
    Taking again the curl of Eq.~(\ref{eqCurlChic}) and computing only its radial component, we get:
\begin{equation}\label{eqCurlCurlChic}
   \left( \frac{n_c}{n_n} \right)' \Delta_{H}  \chi_c =-\hat{r} \cdot \vec{\nabla}\times \vec{\nabla}\times \left(\frac{\vec{f}_{B}}{ \mu n_n} \right).
\end{equation}
Here, $\Delta_{H}$ represents the horizontal Laplacian, which in axial symmetry can be expressed as
\begin{equation}
    \Delta_H \equiv \dfrac{1}{r^{2}\sin \theta}\dfrac{\partial}{\partial \theta}\left(\sin \theta \dfrac{\partial}{\partial \theta}\right),\label{eqDeltaH}
\end{equation}
 and its eigenfunctions are the Legendre polynomials, i.~e.,
\begin{equation}\label{eqll1Pl}
    \Delta_{H} P_l (\cos\theta) = -\dfrac{l(l +1)}{r^{2}} P_l (\cos\theta).    
\end{equation}

Thus, with the aid of equations (\ref{eqCurlCurlChic})--(\ref{eqll1Pl}), we can obtain $\chi_{c,l}$, while $\chi_{n,l}$ is determined by repeating the same procedure but dividing Eq.~(\ref{eq:force_balance_1}) by $\mu n_c$ instead. Thus, for $l \geq 1$, the functions $\chi_{i,l}(r)$ read as
\begin{align}
     \chi_{n,l} &= -\dfrac{r^{2}}{l(l+1)(n_n/n_c)'} \hat{\mathrm{P}}_{l}\left\{\hat{r} \cdot \vec{\nabla}\times \vec{\nabla}\times \left(\frac{\vec{f}_{B}}{\mu n_c} \right)\right\},\label{eq:chinl_geq1}\\
     \chi _{c,l} &=  -\dfrac{r^{2}}{l(l+1)(n_c/n_n)'} \hat{\mathrm{P}}_{l}\left\{\hat{r} \cdot \vec{\nabla}\times \vec{\nabla}\times \left(\frac{\vec{f}_{B}}{\mu n_n} \right)\right\}.\label{eq:chicl_geq1}
\end{align}

\item Next, we obtain the velocity fields. The simplest to obtain is the ambipolar velocity, which is obtained once $X_{n}=\sum_{l\geq 1}\chi_{n,l}(r)P_l(\cos\theta)$ has 
been determined (see Eq.~\ref{eq:v_ambipolar23}). On the other hand, since $n_n \vec{v}_{n}$ is purely solenoidal, its poloidal component can be written as
\begin{equation}
    \vec{v}_{n\text{, Pol}} = \frac{1}{n_n}\vec{\nabla}\times \vec{\nabla}\times (\Psi \hat{r}) \,. \label{eq:v_n23}
\end{equation}
where $\Psi(r, \theta)$ is a scalar field. Taking the radial component of the latter and projecting its $l$'th Legendre component yields 
\begin{equation}
  \hat{\mathrm{P}}_{l}\left\{v_{n,r}\right\} = \frac{l(l+1)\Psi_l}{r^2 n_n} \,,
\end{equation}
where $\Psi_{l}=\hat{\mathrm{P}}_{l}\left\{\Psi\right\}$. This, together with Eq.~\eqref{eq:vnr_anl}, implies
\begin{equation}
    \Psi_{l} =- \dfrac{r^{2}}{l(l+1)(n_c/n_n)'}\hat{\mathrm{P}}_{l}\left\{ \vec{\nabla}\cdot \left(\dfrac{\mu \vec{\nabla} X_n}{\gamma_{nc}}\right)\right\} \,,\label{eq:Psi_l}
\end{equation}
for $l\geq 1$, which can be used to compute $v_{n,\theta}$ from Eq.~\eqref{eq:v_n23}, whereas the term $\Psi_{0}$ is not relevant as it vanishes from the right-hand-side of the latter.
\end{enumerate}

These equations are solved using the pseudo-spectral code \texttt{Dedalus}\footnote{www.dedalus-project.org}: a flexible framework for spectrally solving differential equations, encompassing boundary, initial, and eigenvalue problems \citep{Burns2020_DedalusFlexibleFramework}. It expands in terms of Jacobi polynomials and spherical harmonics in the radial and angular domains, respectively (for more details, see, e.~g., \citealt{Vasil2019_TensorCalculusSpherical}).

To compute the velocity field, we implement the procedure described above.
$\chi_{i,l}$ $(i=n,c)$ for $l \geq 1$ are computed by straightforward substitution in Eqs.~\eqref{eq:chinl_geq1} and \eqref{eq:chicl_geq1}. 
After computing them, obtaining $\vec{v}_{ad}$ and $\vec{v}_{n}$ is a simple substitution in \texttt{Dedalus} (Eqs.~\ref{eq:v_ambipolar23}, \ref{eq:v_n23}, and \ref{eq:Psi_l}).
If needed, we can also determine $\chi_{n,0}$ and $\chi_{c,0}$. To do this, we need to solve a boundary value problem consisting of Eq.~\eqref{eq:Fbalance_r} together with conditions \eqref{eq:cond_chi_n0}-\eqref{eq:cond_chi_c0}, and the integral conditions \eqref{eq:cons_n}-\eqref{eq:cons_c}. In the implementation of this approach, we set the truncation order to $n_r =100$ and $n_{\theta}=71$ radial and angular polynomials, respectively.

Note that the pseudo-spectral approach just described has the same limitations as the explicit approach; namely, it does not provide a way to compute the toroidal component of $\vec{v}_{n}$ and $\vec{v}_{c}$, which makes the approach unsuitable to perform time evolving simulations, and it does not allow to impose boundary conditions at the crust-core interface.

\section{Fictitious Friction Approach}
\label{sec:fictitious_friction}

In Sect.~\ref{sec:semi-anallytical}, we saw how the set of Eqs.~\eqref{eq:continuity_n2}, \eqref{eq:continuity_c2}, \eqref{eq:force_balance_1}, and \eqref{eq:v_ambipolar} can, in principle, be solved for any given magnetic field configuration. However, the numerical implementation is cumbersome, as it requires knowledge of high-order derivatives of the magnetic field \citep{Ofengeim2018_FastMagneticField}.
In \citet{Castillo2020_TwofluidSimulationsMagnetic}, we instead introduced a FF force in Eq.~\eqref{eq:force_balance_1} (see \citealt{Yang1986_ForcefreeMagneticFields, Roumeliotis1994_NumericalStudySudden, Hoyos2008_MagneticFieldEvolution, Hoyos2010_AsymptoticNonlinearSolutions, Vigano2011_ForcefreeTwistedMagnetospheres}), defined as
\begin{equation}
	\vec{f}_{\zeta}=-\zeta n_n\vec{v}_n\,,
\end{equation}
where $\zeta$ is a (small) adjustable parameter, which is a strategy that, to our knowledge, was first described by \citet{Chodura1981_3DCodeMHD}. 
With this additional force, the force balance equation takes the form:
\begin{equation}
	\vec{f}_{B}+\vec{f}_{n}+\vec{f}_{c}-\zeta n_n\vec{v}_n=0, 
	\label{eq:force_balance_2}
\end{equation}
from which the neutron velocity can be determined:
\begin{equation}
	\vec{v}_n=\frac{1}{\zeta n_n}\left(\vec{f}_{B}+\vec{f}_{n}+\vec{f}_{c}\right) 
	\,.
	\label{eq:v_neutrons}
\end{equation}
Within the framework of the FF approach, instead of solving the original system of 
Eqs.~\eqref{eq:continuity_n2}, \eqref{eq:continuity_c2}, \eqref{eq:force_balance_1}, and \eqref{eq:v_ambipolar}, 
it is proposed to solve a modified system consisting of equations 
Eqs.~\eqref{eq:continuity_n2}, \eqref{eq:continuity_c2}, 
\eqref{eq:v_ambipolar}, and \eqref{eq:v_neutrons}. 
This system of equations will hereafter be referred to as the FF equations or FF system, and the corresponding calculation method will be called the FF approach. 
The main advantage of this approach is the simplicity of calculating all three components of the neutron velocity $\vec{v}_n$ 
using Eq.~\eqref{eq:v_neutrons}.
Note that in the limiting case $\zeta\rightarrow\infty$ we recover the one-fluid approximation, in which neutrons are modeled as a motionless background (see, e.g., \citealt{Castillo2017_MagneticFieldEvolution}).

But how valid is the FF approach?  
To answer this question, let us assume that the solution of 
Eqs.~\eqref{eq:continuity_n2}, \eqref{eq:continuity_c2}, \eqref{eq:force_balance_1}, and \eqref{eq:v_ambipolar}
obtained using the rigorous method of Sect.~\ref{sec:semi-anallytical:explicit} is known. 
We will denote the quantities describing this solution with the superscript $^{(0)}$.  
For example, $\chi^{(0)}_{n}$ and $\vec{v}^{(0)}_{n}$ represent the relative neutron chemical potential 
perturbation 
and neutron velocity in the quasi-stationary approximation, respectively, for a given $\vec{B}$.

Now, suppose that for a sufficiently small dimensionless parameter $\zeta$ (in the units specified in Table~\ref{tab:normalization}), 
the solution of the FF system can be expressed as a series expansion in $\zeta$. 
For instance, for the quantities $\chi_{n}$ and $\vec{v}_n$, we can write:
\begin{gather}
	\chi_{n}=\chi^{(0)}_{n}+\zeta\chi^{(1)}_{n}+\zeta^2\chi^{(2)}_{n}+\dots \,,\\
	\vec{v}_n=\vec{v}^{(0)}_{n}+\zeta\vec{v}^{(1)}_{n}+\zeta^2\vec{v}^{(2)}_{n}+\dots \,,
\end{gather}
(Similar expansions are assumed to hold for other problem parameters, such as $\vec{f}_{c}$, $\vec{f}_{n}$, etc.)  
Substituting these expansions into Eq.~\eqref{eq:v_neutrons} gives:
\begin{equation}
	\begin{split}
		\gamma_{cn}n_c n_n[ \zeta\vec{v}^{(0)}_{n} +\zeta^2\vec{v}^{(1)}_{n}+\dots] 
		=\,\, & [\vec{f}_{B}+\vec{f}^{(0)}_{n}+\vec{f}^{(0)}_{c}] + \\
		& \zeta[\vec{f}^{(1)}_{n}+\vec{f}^{(1)}_{c}] 
		+ \dots \,.
	\end{split}
\end{equation}
Equating terms of the same order in $\zeta$ yields an infinite set of equations relating 
quantities of order $k$ and $k+1$ ($k=0,1,\ldots$).  
The first two of these equations are:
\begin{align}
	&
	\vec{f}_{B}+\vec{f}^{(0)}_{n}+\vec{f}^{(0)}_{c}=0 \,,
	&
	\label{eq11}\\
	&
	\vec{v}^{(0)}_{n} = \frac{1}{\gamma_{cn}n_c n_n}\left[\vec{f}^{(1)}_{n}+\vec{f}^{(1)}_{c} \right]\,.
	&
	\label{eq22}
\end{align}
The first equation is satisfied automatically, as it coincides with the force balance equation 
\eqref{eq:force_balance_1} of our exact problem.  
The second equation, in turn, provides a relationship between the neutron velocity $\vec{v}^{(0)}_{n}$ of the exact solution and the first-order corrections $\vec{f}^{(1)}_{n}$ and $\vec{f}^{(1)}_{c}$ to the forces $\vec{f}^{(0)}_{n}$ and $\vec{f}^{(0)}_{c}$.  
It can be shown that the remaining Eqs.~\eqref{eq:continuity_n2}, \eqref{eq:continuity_c2}, and 
\eqref{eq:v_ambipolar} allow, in principle, for the complete calculation of all first-order 
corrections 
in terms of the zeroth-order quantities (computed using the rigorous approach).
The higher-order terms can be found following the same strategy.  

From this example, the essence of the FF approach becomes clear.  
Within this approach, the solution approximately corresponds to the exact solution of  
Eqs.~\eqref{eq:continuity_n2}, \eqref{eq:continuity_c2}, \eqref{eq:force_balance_1}, and \eqref{eq:v_ambipolar}
with an accuracy of $\mathcal{O}(\zeta)$. For instance:
$\vec{v}_n-\vec{v}^{(0)}_{n}\sim\mathcal{O}(\zeta)$.  
Thus, for sufficiently small $\zeta$, the FF approach should yield a solution that deviates 
only slightly from the exact calculation.

We have just discussed how, given the exact solution of the system 
Eqs.~\eqref{eq:continuity_n2}, \eqref{eq:continuity_c2}, \eqref{eq:force_balance_1}, and \eqref{eq:v_ambipolar},  
the solution of the FF system can be derived.  
However, in practice, the FF approach was specifically proposed to avoid solving the exact system 
of Eqs.~\eqref{eq:continuity_n2}--\eqref{eq:v_ambipolar}.  
Suppose we choose a sufficiently small parameter $\zeta$ and find the solution of the FF 
equations.  
Will this solution always be close to the exact solution of the system 
\eqref{eq:continuity_n2}--\eqref{eq:v_ambipolar}?  
Unfortunately, the answer is negative.  
As discussed below, the FF system requires more boundary conditions than the 
original system 
\eqref{eq:continuity_n2}--\eqref{eq:v_ambipolar}.  
Specifically, to uniquely solve the FF equations, some boundary conditions must be imposed, e.~g., specifying the $r$-components of the velocities $\vec{v}_n$ and $\vec{v}_c$ at the crust-core interface.
Meanwhile, in the exact system, these components are determined automatically during the solution 
process, 
as they depend on the magnetic field structure.  
Thus, the FF system of equations describes a broader class of solutions,  
which only becomes close [to an accuracy of $\sim \mathcal{O}(\zeta)$] 
to the exact solution if the boundary conditions for the FF system (values of the $r$-components of 
the velocities $\vec{v}_n$  
and $\vec{v}_c$ at the boundary) are chosen to be sufficiently close to the corresponding boundary 
values obtained in the exact calculation.  
This interpretation has been verified numerically; see Sect.~\ref{sec:results:comparison:FF} for details.  
With these caveats, the FF approach can be successfully used to determine velocity fields for a 
given magnetic field configuration, as described below.

To solve the FF system for a given initial condition for the magnetic field, we can replace the velocities from Eqs.~\eqref{eq:v_ambipolar} and \eqref{eq:v_neutrons} into the continuity Eqs.~\eqref{eq:continuity_n2} and \eqref{eq:continuity_c2}, together with Eqs.~\eqref{eq:neutron force} and \eqref{eq:charged force}. After rearranging terms, leaving only the terms explicitly dependent on the magnetic field on the right-hand side of the equations, we obtain
\begin{gather}
	\vec{\nabla}\cdot\left(n_n\mu\vec{\nabla} \chi_n+n_c\mu\vec{\nabla} \chi_c\right)
 = \vec{\nabla}\cdot\vec{f}_{B}\label{eq:continuity_n4}\,,\\
	\vec{\nabla}\cdot\left[n_c\mu\vec{\nabla} \chi_n +  g(\zeta)n_c\mu\vec{\nabla} \chi_c    \right]
  = \vec{\nabla}\cdot\left[ g(\zeta) \vec{f}_{B}\right]  \,. \label{eq:continuity_c4}
\end{gather}
where we have introduced a dimensionless radial function 
\begin{equation}
    g(\zeta,r)= \frac{\zeta}{\gamma_{cn} n_n} + \frac{n_c}{n_n}\,. \label{eq:gz}
\end{equation}
The differential equations \eqref{eq:continuity_n4}, \eqref{eq:continuity_c4}, and integral conditions \eqref{eq:cons_n} and \eqref{eq:cons_c} fully define the functions $\chi_c$ and $\chi_n$ when combined with appropriate boundary conditions such as the \textit{non-penetration} conditions at the crust-core interface for the velocity fields, i.~e., $\vec{v}_{n,r} = \vec{v}_{ad,r} = 0$, which translates into
\begin{gather}
    \left.\frac{\partial \chi_c}{\partial r}\right |_{r=R} = \left.\frac{f_{B,r}}{n_c \mu}\right |_{r=R}\,, \label{eq:boudary1}\\
    \left.\frac{\partial \chi_n}{\partial r}\right |_{r=R} =0\,. \label{eq:boudary2}
\end{gather}
We choose these boundary conditions as they can be easily implemented numerically.
However, it should be noted that, as shown by \citet{Ofengeim2018_FastMagneticField}, for an arbitrary magnetic field, the exact solution to the set of Eqs.~\eqref{eq:continuity_n2}, \eqref{eq:continuity_c2}, \eqref{eq:force_balance_1}, and \eqref{eq:v_ambipolar} will typically yield non-null radial components of the velocities at the boundary.

\subsection{Numerical implementation using finite difference/finite volume}
\label{sec:numerical:FD}

In order to numerically solve our equations using a finite difference/finite volume implementation, we discretize the relevant variables over a staggered polar grid composed of $N_r$ points that can be homogeneously or inhomogeneously distributed in the radial direction and $N_\theta$ points equally spaced in the polar direction, following the rule
\begin{gather}
r_i=\left(\frac{i-1}{N_r-1}\right)^u\quad i=1,\dots, N_r \,, \label{eq:r_i}\\
\theta_j=\pi \frac{j-1}{N_\theta-1}=(j-1)\Delta\theta\quad j=1,\dots, N_\theta \,.
\end{gather}
The exponent $u$ will be specified later. We discretize the relevant variables on this grid in the following way, where half-integer indices denote averages, e.~g., $r_{i+1/2}\equiv(r_{i}+r_{i+1})/2$:
\begin{enumerate}
    \item $\alpha$ is evaluated at the corner of each cell, e.~g., $\alpha_{i,j}=\alpha(r_{i}, \theta_{j})$.
    \item $\beta$, $\chi_n$, $\chi_c$, and the $\phi$ component of the forces and velocities are evaluated at the center of each cell, e.~g., $\beta_{i,j}=\beta(r_{i+1/2}, \theta_{j+1/2})$.
    \item The radial components of vector fields, such as $\vec{B}$, forces, and velocities are evaluated at the centers of cell faces defined by constant $r=r_i$, e.~g., $B_{r\,i,j}=B_r(r_{i}, \theta_{j+1/2})$.
    \item The polar components of vector fields, such as $\vec{B}$, forces, and velocities are evaluated at the centers of cell faces defined by constant $\theta=\theta_j$, e.~g., $B_{\theta\,i,j}=B_\theta(r_{i+1/2}, \theta_{j})$.
\end{enumerate}

The value of the divergence of a vector field $\vec{F}$ at the center of each cell $(\vec{\nabla}\cdot\vec{F})_{i,j} =\left.\vec{\nabla}\cdot \vec{F}\right|_{r_{i+1/2},\theta_{j+1/2}}$ is given by
\begin{equation}
\begin{split}
(\vec{\nabla}\cdot \vec F)_{i,j} = \frac{1}{\mathcal{V}_{i,j}}\big(& S_{r\,i+1,j}F_{r\,i+1,j}-S_{r\,i,j}F_{r\,i,j} \\
&+S_{\theta\,i,j+1}F_{\theta\,i,j+1}-S_{\theta\,i,j}F_{\theta\,i,j}\big) \,,
\end{split}
\end{equation}
where $F_{r\,i,j} = F_r(r_i,\theta_{j+1/2})$, $F_{\theta\,i,j} = F_\theta(r_{i+1/2},\theta_{j})$; $S_{r\,i,j}$ and $S_{\theta\,i,j}$ are the surface areas of the faces of each cell at constant $r=r_i$ and $\theta=\theta_j$, respectively; and $\mathcal{V}_{i,j}$ is the volume of the cell.
Therefore,
\begin{equation}
\begin{split} \label{eq:divnumerica}
(\vec{\nabla}\cdot \vec{F})_{i,j} =&\, 3\frac{r_{i+1}^2F_{r\,i+1,j}-r_{i}^2F_{r\,i,j}}{r_{i+1}^3-r_{i}^3} \\
&+\frac{3(r_{i+1}^2-r_{i}^2)}{2(r_{i+1}^3-r_{i}^3)}\frac{\sin\theta_{j+1}F_{\theta\,i,j+1}-\sin\theta_{j}F_{\theta\,i,j}}{\cos\theta_j-\cos\theta_{j+1}} \,.
\end{split}
\end{equation}

Assuming boundary conditions $v_{n,r}(r=R)=v_{c,r}(r=R)=0$, Eqs.~\eqref{eq:continuity_n4} and \eqref{eq:continuity_c4} can be solved for the relative chemical perturbations $\chi_n$ and $\chi_c$.
Considering all grid cells, we have two sets of variables, $\{\chi_{n\, i,j}\}_{i=1,j=1}^{N_r-1,N_\theta-1}$ and $\{\chi_{c\, i,j}\}_{i=1,j=1}^{N_r-1,N_\theta-1}$, giving a total of $2(N_r-1)(N_\theta-1)$ unknowns.
Eqs.~\eqref{eq:continuity_n4} and \eqref{eq:continuity_c4} are evaluated at the center of each grid cell. Thus, each of them provides $(N_r-1)(N_\theta-1)$ equations, making a total of $2(N_r-1)(N_\theta-1)$. 
Note that these equations only involve spatial derivatives of the relative chemical perturbations.
Thus, they determine $\chi_n$ and $\chi_c$ up to additive constants. Fixing these constants requires extra conditions provided by Eqs.~\eqref{eq:cons_n} and \eqref{eq:cons_c}.
Therefore, for the system of equations to be consistent, we must replace two redundant equations from the discretization of Eqs.~\eqref{eq:continuity_n4} and \eqref{eq:continuity_c4}. We choose to replace the equations corresponding to the grid cell $(i,j)=(1,1)$.

In the evaluation of each divergence, the boundary conditions on the velocities must be taken into account. 
For each cell touching the symmetry axis, we impose the fluxes through the corresponding face of the cell to be null (i.~e., $\theta=0$ or $\theta=\pi$).
In the cells touching the center of the star, we impose fluxes through the face defined by $r=0$ to be null, and at the faces defined by $r=1$ we impose $v_{n,r}=0$ and $v_{ad,r}=0$, thus fluxes through those faces are null as well. In other words, radial fluxes at $r=1$ of the cells touching the boundary are equal on both sides of equations \eqref{eq:continuity_n4} and \eqref{eq:continuity_c4}.

Schematically, the linear system described before can be represented by a vector of unknown values of the relative chemical potential perturbations $\vec{\chi}=\left\{{ \{\chi_{n\, i,j}\}_{i=1,j=1}^{N_r-1,N_\theta-1} , \{\chi_{n\, i,j}\}_{i=1,j=1}^{N_r-1,N_\theta-1}}\right\}$, a vector $\vec{b}$, containing the right-hand side of Eqs.~\eqref{eq:continuity_n4}, \eqref{eq:continuity_c4}, \eqref{eq:cons_n} and \eqref{eq:cons_c}, and a sparse matrix $\Lambda$, generated from the coefficients extracted from the discretization of the left-hand side of the previous equation. 
\begin{equation}\label{eq:Lambda LS}
\Lambda\begin{Bmatrix}
    \{\chi_{n\, i,j}\}_{i=1,j=1}^{N_r-1,N_\theta-1}\\
    \{\chi_{c\, i,j}\}_{i=1,j=1}^{N_r-1,N_\theta-1}
  \end{Bmatrix}
  =\vec{b}\equiv
  \begin{Bmatrix}
    \{\text{rhs\eqref{eq:continuity_n4}}\}_{ (i,j)\neq (1,1) }^{N_r-1,N_\theta-1}\\
    \{\text{rhs\eqref{eq:continuity_c4}}\}_{ (i,j)\neq (1,1) }^{N_r-1,N_\theta-1}\\ 
    0\\ 
    0
  \end{Bmatrix}\,.
\end{equation}
Note that all the dependence on the magnetic field is in $\vec{b}$. Thus, $\Lambda$ needs to be inverted only once in a time-evolving simulation. It should also be noted that this approach has some limitations. Setting a very small value of $\zeta$, is not only limited by the integration time. As $\zeta$ decreases, the matrix $\Lambda$ becomes progressively more singular. 

In summary, this solution method consists of the following steps:
\begin{enumerate}
\item We fix an initial seed for the magnetic field.
\item From the magnetic field, we compute $\vec{b}$, and use it to solve for the chemical potential perturbations:
\begin{equation}
\vec{\chi}=\Lambda^{-1}\vec{b}\,.\label{eq:chi solve}
\end{equation}
\item We replace the values of $\chi_n$ and $\chi_c$ into Eqs.~\eqref{eq:v_ambipolar} and \eqref{eq:v_neutrons}, obtaining the relevant velocities.
\end{enumerate}
In case we are also interested in following the time evolution:
\begin{enumerate}\addtocounter{enumi}{3}
\item We compute the time derivative of the magnetic field from Eqs.~\eqref{eq:faraday} and use it to evolve the magnetic field one time-step (see more details in Sect.~\ref{sec:evolution}).
\item We iterate over this procedure, going back to step 1, now using the evolved magnetic field.
\end{enumerate}

\subsection{Numerical implementation with the \texttt{Dedalus} pseudo-spectral code}

As an alternative numerical implementation of the FF approach, we solved Eqs.~\eqref{eq:continuity_n4} and \eqref{eq:continuity_c4} with the normalization described in Table~\ref{tab:normalization} using \texttt{Dedalus}. The truncation orders were set to 
$n_r =92$ for the radial polynomials and 
$n_\theta = 65$ for the angular polynomials. As explained in Sect.~\ref{sec:semi-anallytical:spectral}, \texttt{Dedalus} expands the radial and angular domains using Jacobi polynomials and spherical harmonics, respectively, leveraging their properties to solve the system in Eq.~\eqref{eq:chi solve} efficiently.

We remark that all regularity conditions imposed by axial symmetry are naturally satisfied, as the polynomial basis used in the expansion is inherently regular at the axis and the origin. Additionally, the boundary conditions, Eqs.~\eqref{eq:boudary1} and \eqref{eq:boudary2}, are enforced in \texttt{Dedalus} using the $\tau$-method (see e.~g., \citealt{Boyd2001_ChebyshevFourierSpectral}).

\section{Background neutron star model}
\label{sec:EoS}

\begin{figure}
	\centering
	\includegraphics[width=\linewidth]{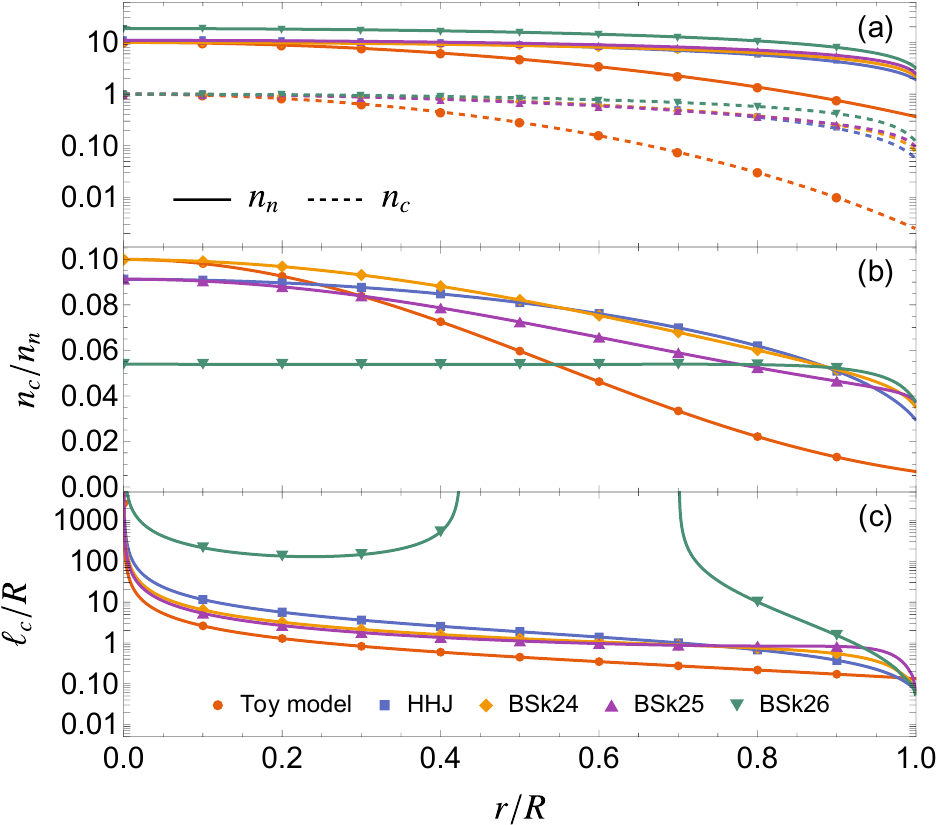}
	\caption{Comparison between the five different equations of state studied. (a) $n_n$ and $n_c$, in units of $n_{c}(r=0)$; (b) the ratio $n_c/n_n$; and (c) $\ell_c/R$, where $\ell_{c}\equiv -\left[d\ln(n_{c}/n_{n})/dr\right]^{-1}$ (see Sect.~\ref{sec:evolution:timescales}) and $R$ is the core radius; all as functions of the normalized radial coordinate, $r/R$.}
	\label{fig:n-ratio}
\end{figure}

\begin{table}
    \setlength{\tabcolsep}{1pt}
	\caption{Stellar parameters for the five different EoSs studied. The toy model corresponds to a star with a mass of $1.58 M_{\odot}$, while the others have a mass of $1.4 M_{\odot}$.}
	\label{tab:stellar parameters}
	\centering
	\begin{tabular}{lccccc} 
		\hline\hline
  		Name & Star  & Core & $n_0$                      & $\mu_0$     & $\gamma_0$                              \\
  		& radius  & radius, $R$ & & & \\
		Unit & (km)  & (km)  & ($10^{37}\text{cm}^{-3} $) & (MeV)      & ($10^{-46}\text{cm}^3 \text{g s}^{-1}$) \\
		\hline
		Toy model &  8.2 &  6.87   &  60.4    &  1597 &  0.032 \\
		HHJ &  12.17 &  11.2 &  4.23    &  1173 & 1.13\\
		BSk24 &  12.65 &  11.6  &  3.62    &  1140 & 1.58\\
		BSk25 &  12.43 &  11.4 &  3.41    &  1141 & 1.52\\
  		BSk26 &  11.8 &  10.9 &  2.58    &  1173 & 1.10\\
		\hline
	\end{tabular}
\end{table}

We perform our analysis considering five different EoSs. One is the same toy model described in \citet{Castillo2020_TwofluidSimulationsMagnetic}. There, neutrons and protons are treated as a non-relativistic Fermi gas, and electrons are treated as an extremely relativistic Fermi gas. Both neutrons and protons are considered to have the same mass $m$.
Therefore, the chemical potentials are related to the particle densities by 
\begin{gather}
\mu_n(r) = m c^2 + \frac{p_{Fn}(r)^2}{2m}\,,\\
\mu_c(r) = m c^2 + \frac{p_{Fc}(r)^2}{2m} + p_{Fc}(r)c \,,
\end{gather}
where $p_{Fi}(r) = \hbar\left[3\pi^2n_i(r)\right]^{1/3}\,$ ($i=n,c$) are the Fermi momenta of neutrons and charged particles, respectively.
Thus, the density profiles can be written as
\begin{gather}
n_n(r)=\left[\frac{2mc}{\hbar(3\pi^2)^{1/3}}n_c(r)^{1/3} + n_c(r)^{2/3}\right]^{3/2}\,,\label{eq:chemical-equilibrium-n}\\
n_c(r)=n_{0}\,\text{sinc}^6(0.7\pi r/R)\,,\label{eq:nc_A}
\end{gather}
where $\text{sinc}(x)=\sin(x)/x$. Eq.~\eqref{eq:chemical-equilibrium-n} comes from the chemical equilibrium condition, $\mu_n(r)=\mu_c(r)$, and Eq.~\eqref{eq:nc_A} is an analytic approximation to the radial profile obtained by numerically integrating the (non-relativistic) stellar structure equations.

The other EoSs used are HHJ \citep{Heiselberg1999_PhaseTransitionsNeutron}, BSk24, BSk25, and BSk26 \citep{Goriely2013_FurtherExplorationsSkyrmeHartreeFockBogoliubov},
which have been adapted for $npe$ matter.
For all realistic models the central density is adjusted so that the mass of the star is $1.4 M_{\odot}$.
The basic parameters of the stellar models are summarized in Table~\ref{tab:stellar parameters}. 
Unlike the toy model, these realistic EoSs consider strong interactions by means of a coefficient $K\neq 0$ (see Sect.~\ref{sec:model}). A comparison between $n_n$, $n_c$, and  $n_c/n_n$ as functions of $r$ for the different models can be seen in Figs.~\ref{fig:n-ratio}(a) and (b).
The collision coefficient $\gamma_{cn}$ between charged particles and neutrons for the different models is computed from $J_{pn}(r,T)=n_c(r)n_n(r)\gamma_{cn}(r,T)$ given by \citet{Yakovlev1991_ElectricalConductivityNeutron}, where $T$ is the temperature.

\section{Analytical magnetic field models}
\label{sec:magnetic_models}

In our simulations, initial conditions are given by analytical expressions for $\alpha$ and $\beta$ as functions of $r$ and $\theta$. These functions are normalized so that the rms value of the magnetic field in the core is 1: 
\begin{equation}
    \langle \vec{B} \rangle\equiv\left(\frac{1}{{\cal V}_c}\int_{{\cal V}_c}\vec{B}^2d{\cal V}\right)^{1/2}=1\,.
\end{equation}
We will focus on the following models:
\begin{itemize}
\item Model I:

Arguably, the simplest expressions for a purely poloidal magnetic configuration are of the form 
\begin{equation}
    \alpha(r,\theta)=f(r)\sin^2\theta\,,
\end{equation}
where $f(r)$ is a polynomial on $r$. In particular, the following has been widely used in the literature \citep{Akgun2013_StabilityMagneticFields, Armaza2015_MAGNETICEQUILIBRIABAROTROPIC, Passamonti2017_RelevanceAmbipolarDiffusion, Ofengeim2018_FastMagneticField, Castillo2020_TwofluidSimulationsMagnetic}:
\begin{equation}
    \alpha_{\text{I}}(r,\theta)=\frac{35}{8}\sqrt{\frac{11}{118}} \left( r^2 - \frac{6}{5} r^4 + \frac{3}{7} r^6 \right)\sin^2\theta\,.
\end{equation}
The numerical coefficients on the polynomial come from imposing continuity of $\vec{B}$ on the surface of the core with a current-free dipole solution outside [$f(r) \propto 1/r$], continuity of the toroidal component of the current density $J_\phi$ (although this is not a physical requirement), and the normalization of the magnetic field (in addition to a regularity condition at the axis, which does not allow terms proportional to $r^0$ and $r^1$).\\

\item Model II, a model with null radial velocities at the crust-core interface:

Following \citet{Ofengeim2018_FastMagneticField}, we can construct poloidal magnetic field configurations for which the radial components of both $\vec{v}_{n}$ and $\vec{v}_{ad}$ vanish at the crust-core interface.
First, we take $\alpha$ as a multipolar expansion of the form
\begin{equation}
    \alpha(r,\theta)=\sum_{l\geq 1} C_l\alpha_l(r,\theta)\,,
\end{equation}
where
\begin{equation}
  \alpha_l(r,\theta)=f_l(r)P_l^1(\cos\theta)\sin\theta \,, 
\end{equation}
$C_l$ are constants that determine the relative weights of each multipole, $f_l(r)=\sum_i a_{l,i}r^i $ are polynomials of $r$, and $P_l^1$ are the associated Legendre functions. The potentials $\alpha_l$ are normalized so their generated magnetic field $\vec{B}_l\equiv\vec{\nabla}\alpha_l\times\vec{\nabla}\phi$ satisfies $\langle \vec{B}_l \rangle=1$, implying $\sum_l C_l^2=1$. 
For this model, we only consider a dipole; however, equations will be written for a more general case, as they also apply to model III.
The order of the polynomial(s) depends on the number of conditions to be imposed.
We enforce the same three conditions as for model I, namely 
\begin{gather}
f'_l(1)+l\,f_l(1)=0\,,\label{eq:cond B}\\
f''_l(1)-l(l+1)\,f_l(1)=0\,, \label{eq:cond J}\\
\langle \vec{B}_l \rangle=1 \label{eq:cond norm}\,.
\end{gather}
These are, respectively, continuity of $\vec{B}$ at the crust-core boundary considering a current-free field everywhere outside the core; continuity of  $J_\phi$, which is null outside; and a normalization. 

Additional conditions follow from setting $v_{n,r}(1,\theta)=v_{ad,r}(1,\theta)=0$.
If we expand $v_{ad,r}(1,\theta)$, using Eq.~\eqref{eq:v_ambipolar23}, into Legendre polynomials, all Legendre components should be null, which implies $\hat{\mathrm{P}}_{l}\{\partial X_n/\partial r|_{r=1}\}=0$, $l=1,2,\dots$. 
Thus, the first extra condition is
\begin{equation}
    \int_0^\pi \left.\frac{\partial X_n}{\partial r}\right|_{r=1}P_l(\cos\theta)\sin\theta=0 \,,\quad l=1,2,.. \label{eq:condition_vad2}
\end{equation}
Analogously, Eq.~\eqref{eq:vnr_anl} implies $\hat{\mathrm{P}}_{l}\{\vec{\nabla}\cdot[(\mu/\gamma_{cn})\vec{\nabla} X_n] |_{r=1}\}=0$, $l=1,2,\dots$, which, together with Eq.~\eqref{eq:condition_vad2}, yields the second extra condition
\begin{equation}
    \int_0^\pi \left[\frac{\partial^2X_n}{\partial r^2}-\frac{l(l+1)}{r^2}X_n\right]_{r=1}P_l(\cos\theta)\sin\theta=0 \,,\quad l=1,2,... \label{eq:condition_vn2}
\end{equation}
To apply these two conditions, we first need to obtain $X_n$ in terms of the set of coefficients $\{a_{l,i}\}$ in $f_l(r)$. This is done first by computing the magnetic force in terms of the $a_{l,i}$ coefficients, then using \eqref{eq:epsilon_sol}, \eqref{eq:delta_sol}, and \eqref{eq:Xn_ed}.
The number of non-zero Legendre components in $X_n$, and consequently in Eqs.~\eqref{eq:condition_vad2} and \eqref{eq:condition_vn2}, increases nonlinearly with the number of non-zero Legendre components in $\vec{B}$ (as $X_n$ is nonlinear in $\vec{B}$).
For a dipolar magnetic field configuration (corresponding to $l=1$), the only non-zero Legendre component in $X_n$ is $l=2$, which imposes non-linear relations for the coefficients $\{a_{l=1,i}\}$ in Eqs.~\eqref{eq:condition_vad2} and \eqref{eq:condition_vn2}.
We have a total of five conditions; thus, we use a polynomial of five coefficients of the form
\begin{equation}
f_1(r)=a_{1,2}r^2+a_{1,4}r^4+a_{1,6}r^6+a_{1,8}r^8+a_{1,10}r^{10} \,.\label{eq:polynomial}
\end{equation}
Solving for the coefficients, using the conditions previously described, yields eight solutions, four of which, evaluated with the HHJ EoS, are summarized in Table~\ref{tab:coef model ii} and displayed in Fig.~\ref{fig:model II solutions}. The other four are the same, just with a global minus sign. We will focus on ``Solution 1'' (hereafter model II, $\alpha_{\text{II}}$), as in the others, the magnetic flux is fully enclosed inside the core.\\

\begin{table}
    \setlength{\tabcolsep}{4pt}
	\caption{Coefficients for the polynomial in Eq.~\eqref{eq:polynomial}, yielding $v_{n,r}=v_{ad,r}=0$ at the crust-core interface for our background NS model with the HHJ EoS.}
	\label{tab:coef model ii}
	\centering
	\begin{tabular}{lccccc}
		\hline\hline
		 & $a_{1,2}$ & $a_{1,4}$ & $a_{1,6}$ & $a_{1,8}$ & $a_{1,10}$\\
		\hline
		Solution 1 & 2.0454 & -5.1851 & 5.7957  & -3.1009 & 0.64796\\
		Solution 2 & 2.4580 & -8.5909 & 11.025 & -6.1085 & 1.2168\\
		Solution 3 & 1.1947 & 0.35944 & -8.2463 & 10.636 & -3.9434\\
        Solution 4 & 2.0205 & -5.2131 & 3.5166 & 0.52437 & -0.84828\\
		\hline
	\end{tabular}
\end{table}

\begin{figure}
	\centering
	\includegraphics[width=\linewidth]{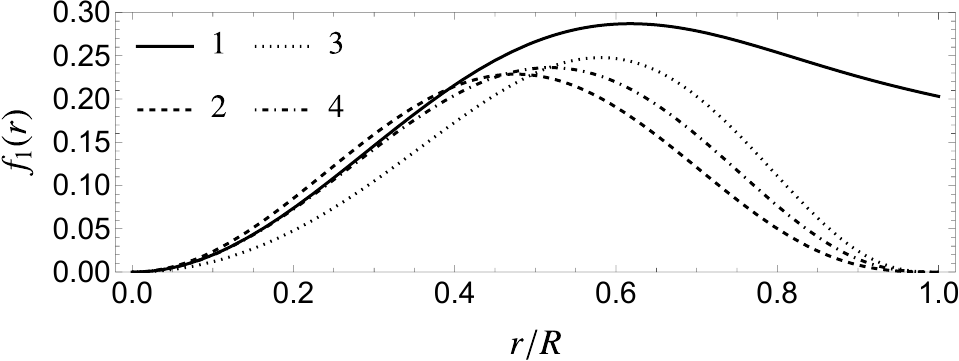}
	\caption{Radial functions $f_1(r)$ for model II. The coefficients for each solution are given in Table~\ref{tab:coef model ii}.
 }
	\label{fig:model II solutions}
\end{figure}

\item Model III, a non-equatorially symmetric model with null radial velocities at the crust-core interface:

Using the conditions from the previous model, we can build a non-equatorially symmetric poloidal magnetic configuration with null radial velocities at the crust-core interface.
For this, we consider a combination of a dipolar and quadrupolar magnetic field configuration of the form $\alpha_{\text{III}}(r,\theta)=\sqrt{0.5}\alpha_1(r,\theta)+\sqrt{0.5}\alpha_2(r,\theta)$, where
\begin{gather}
\alpha_1(r,\theta)=\sum_{i=1}^6 a_{1,2i}r^{2i}\sin^2\theta\\
\alpha_2(r,\theta)=\sum_{i=1}^6 a_{2,2i+1}r^{2i+1}\sin^2\theta\cos\theta
\end{gather}
The number of $a_{l,i}$ coefficients considered comes from the number of conditions needed to fix the configuration. For each of the multipoles ($l=1$ and $l=2$), we have to satisfy the conditions from Eqs.~\eqref{eq:cond B} and \eqref{eq:cond norm}; a total of four. Note that we relaxed condition \eqref{eq:cond J}, as it is not strictly needed. Also, $\alpha_{\text{III}}$ yields a function $X_n(r,\theta)$ that includes non-zero multipoles $l=1,2,3$, and 4. Thus, Eqs.~\eqref{eq:condition_vad2} and \eqref{eq:condition_vn2} yield 8 equations. So, we have set enough coefficients to satisfy the 12 conditions needed for $\alpha_{\text{III}}$. There is a very large number of solutions for this system. The coefficients for the solution we use in this work are presented in Table~\ref{tab:coef model iii}.\\

\begin{table}
	\caption{Coefficients of the polynomials defined for the non-equatorially-symmetric poloidal magnetic configuration $\alpha_{\text{III}}$, yielding $v_{n,r}=v_{ad,r}=0$ at the crust-core interface for our background NS model with the HHJ EoS.}
	\label{tab:coef model iii}
	\centering
	\begin{tabular}{cccccc}
		\hline\hline
		 $a_{1,2}$ & $a_{1,4}$ & $a_{1,6}$ & $a_{1,8}$ & $a_{1,10} $& $a_{1,12}$ \\
         \hline
		-1.8958  & 14.400 & -32.078 & 33.674 & -17.428 & 3.6057 \\
		\hline\hline
		 $a_{2,3}$ & $a_{2,5}$ & $a_{2,7}$ & $a_{2,9} $& $a_{2,11}$ & $a_{2,13}$\\
         \hline
        8.6212  & -30.790 & 47.895 & -39.187 & 16.579  & -2.8737 \\
		\hline
	\end{tabular}
\end{table}

\item Model IV, a non-equatorially symmetric model with a toroidal component:

We use the same non-equatorially symmetric magnetic field introduced in \citet{Castillo2020_TwofluidSimulationsMagnetic}, with poloidal and toroidal components produced by the potentials
\begin{gather}
\alpha_{\text{IV}}(r,\theta) = \sqrt{0.18}\alpha_{\text{I}}(r,\theta)+\sqrt{0.42}\alpha_\text{aux}(r,\theta)\,,\label{eq:alpha3}\\
\beta_{\text{IV}}(r,\theta) = \sqrt{0.4}\beta_\text{aux}(r,\theta) \label{eq:beta3} \,,
\end{gather}
where
\begin{gather}
\alpha_\text{aux}(r,\theta)=3.718\, \left( r^3 -\frac {10}7r^5 +\frac 59r^7 \right)\sin^2\theta\cos\theta\,,\\
\beta_\text{aux}(r,\theta)=112.546\, r^5(1 -r)^2\sin^2\theta \sin(\theta-\pi/5)\,.
\end{gather}
The quadrupolar poloidal potential $\alpha_\text{aux}$ (which satisfies the same conditions as $\alpha_{\text{I}}$), and the toroidal potential $\beta_\text{aux}$ are both normalized so $\langle \vec{B}_{\text{Pol}}^\text{aux} \rangle=1$ and $\langle \vec{B}_{\text{Tor}}^\text{aux} \rangle=1$, respectively\footnote{The prefactor of $\alpha_\text{aux}$ given in \citet{Castillo2020_TwofluidSimulationsMagnetic} is not correct.}; so $\alpha_{\text{IV}}$ and $\beta_{\text{IV}}$ will produce a magnetic configuration in which the toroidal magnetic field has 40\% of the internal magnetic energy.

\end{itemize}

\begin{figure}
	\centering
	\includegraphics[width=0.8\linewidth]{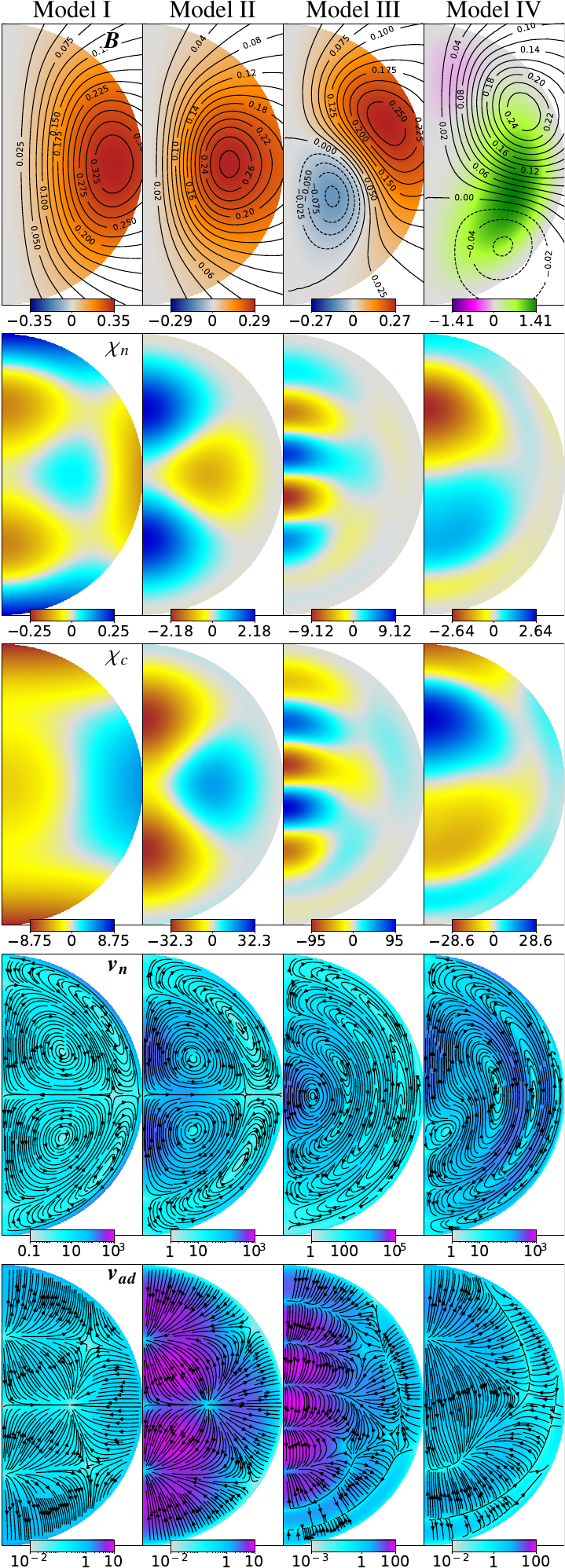}
	\caption{Magnetic field configurations, relative chemical potential perturbations, and velocity fields produced by the models I, II, III, and IV. For $\vec{B}$, lines represent the poloidal magnetic field lines labeled by the corresponding value of $\alpha$. For models I--III, color represents the poloidal potential $\alpha$, while in model IV, it represents the toroidal potential $\beta$. For $\chi_n$ and $\chi_c$, color represents their local value. For $\vec{v}_{n}$ and $\vec{v}_{ad}$, lines represent the poloidal component and color its magnitude; these are computed through the explicit semi-analytic approach described in Sect.~\ref{sec:semi-anallytical:explicit} using the background NS model with the HHJ EoS.
 }
	\label{fig:models}
\end{figure}

The magnetic field models I, II, III, and IV are shown in the first row of Fig.~\ref{fig:models}.

\section{Time evolution}
\label{sec:evolution}

The time-evolving equations for the magnetic potentials can be obtained from Eq.~\eqref{eq:faraday}, together with Eq.~\eqref{eq:BAlfaBeta}
\begin{gather}
\frac{\partial\alpha}{\partial t}=r\sin\theta\left[\left(\vec{v}_n+\vec{v}_{ad}\right)\times\vec{B}\right]\cdot\hat\phi\,, \label{eq:evolucionalfa}\\
\frac{\partial\beta}{\partial t}=r^2\sin^2\theta\vec{\nabla}\cdot\left\lbrace\frac{\left[\left(\vec{v}_n+\vec{v}_{ad}\right)\times\vec{B} \right]\times\hat{\phi}}{r\sin\theta} \right\rbrace\,. \label{eq:evolucionbeta}
\end{gather}

The evolution in the core and crust of the star are coupled.
For simplicity, in this work we assume that the crust is a perfect resistor ($\vec{\nabla}\times\vec {B}=0$). Therefore, the magnetic field outside of the core can be written as $\vec {B}=\vec{\nabla}\varphi$, where $\varphi(r,\theta)=\sum_{l=1}^{\infty}(b_l/r^{l+1})P_l (\cos\theta)$ and $b_l$ are coefficients.
The magnetic field in the crust and outside of the star is entirely determined by the radial component of the magnetic field at the crust-core interface, where continuity at that region implies 
$b_l=-R^{l+2}\hat{\mathrm{P}}_{l} \{B_r(R,\theta)\}/(l+1)$ (see \citealt{Marchant2011_RevisitingFlowersRudermanInstability}). Of course, in practice, we are only considering a finite number $N_{\text{Exp}}$ of external multipoles. 

\subsection{Hydromagnetic quasi-equilibrium}
\label{sec:evolution:GS}

As discussed in Sect.~\ref{sec:intro}, after a few Alfvén times, the magnetic field will evolve towards a hydromagnetic quasi-equilibrium state in which, under axial symmetry, the toroidal magnetic force must be null, since the fluid forces $\vec{f}_i=-n_i\mu\vec{\nabla}\chi_i$ ($i=n,c$), which are purely poloidal, cannot balance it.
This imposes a strong condition on the magnetic field; Eq.~\eqref{eq:BAlfaBeta} implies $f_{B,\phi}\propto |\vec{\nabla}\alpha\times\vec{\nabla}\beta| = 0$; which can only be achieved if there is a relation between the magnetic potentials [e.~g., $\beta=\beta(\alpha)$]. As in our model there is no toroidal field outside of the core, the latter implies that the toroidal field must be confined to the regions in which the poloidal magnetic field lines close inside of the core, forming a ``twisted torus'' configuration \citep{Chandrasekhar1956_EquilibriumMagneticStars, Braithwaite2004_FossilOriginMagnetic}. It also implies that the magnetic force takes the form
\begin{equation}
    \vec{f}_B=-\frac{\Delta^*\alpha+\beta d\beta/d\alpha}{4\pi r^2\sin^2\theta}\vec{\nabla}\alpha\,,
    \label{eq:magnetic_force}
\end{equation}
where 
\begin{equation} 
\Delta^*  \equiv \frac{\partial^2}{\partial r^2} + \frac{\sin\theta}{r^2} \frac{\partial}{\partial\theta}\left( \frac{1}{\sin\theta}\frac{\partial}{\partial\theta}\right) \label{eq:shafranov_op}
\end{equation}
is the so-called ``Grad-Shafranov (GS) operator.'' 

As time passes, this quasi-equilibrium is slowly eroded as the magnetic field lines, which are frozen into the charged particle fluid, keep moving relative to the neutron fluid with velocity $\vec{v}_{ad}$. Eventually, a final {\it barotropic} equilibrium state is reached on a timescale $\sim t_B$ (see Sect.~\ref{sec:evolution:timescales}), in which the magnetic force is mainly balanced by the pressure gradient and gravitational force of the charged particle fluid, while neutrons stay in diffusive equilibrium,
\begin{gather}
\vec{f}_{B}-n_c\mu\vec{\nabla}\chi_c = 0\,, \label{eq:GS_1}\\
\vec{\nabla}\chi_n=0 \,. \label{eq:GS_2}
\end{gather}
The latter formula indeed means diffusive equilibrium, since it is equivalent to
the condition
$\nabla \delta \mu_n+(\delta \mu_n/c^2) \nabla\Phi=0$ (see Eq.\ \ref{eq:hydrostatic equilibrium} and the definition of the function $\chi_n$).

Eqs.~\eqref{eq:magnetic_force} and \eqref{eq:GS_1} imply that, in this kind of equilibrium, not only $\beta$, but also $\chi_c$, must be a function of $\alpha$, and these variables must satisfy the so-called ``GS equation''
\begin{equation}
\Delta^*\alpha+\beta\frac{d\beta}{d\alpha} + 4\pi r^2\sin^2\theta\, n_c\mu\frac{d\chi_c}{d\alpha} = 0 \,.\label{eq:GS}
\end{equation}
Solutions to this equation are called ``GS equilibria'' \citep{Grad1958_HydromagneticEquilibriaForcefree,Shafranov1966_PlasmaEquilibriumMagnetic} and have been widely studied in the literature (e.~g., \citealt{Armaza2015_MAGNETICEQUILIBRIABAROTROPIC} and references therein).
Note that in this equilibrium $\chi_n$ must be uniform, but $\chi_c$ is non-uniform. Together with Eq.~\eqref{eq:dmun}, this implies that, unlike what one might have expected, $\delta n_n$ cannot be uniform.

\subsection{Time-scale estimates}
\label{sec:evolution:timescales}

From Eq.~\eqref{eq:faraday}, we can estimate the timescale for magnetic evolution in the core of the star as 
\begin{equation}
t_B\sim\frac{R}{|\vec{v}_c|}=\frac{R}{|\vec{v}_{n}+\vec{v}_{ad}|}\,,
\end{equation}
The usual assumptions found in the literature  \citep{Goldreich1992_MagneticFieldDecay,Thompson1995_SoftGammaRepeaters,Castillo2017_MagneticFieldEvolution,Passamonti2017_RelevanceAmbipolarDiffusion} are that the neutron velocity can be neglected and that the magnitude of the poloidal component of the ambipolar velocity can be estimated from Eq.~\eqref{eq:v_ambipolar} as $v_{ad\text{, Pol}}\sim f_B/(\gamma_{cn}n_c n_n)$ ($v_{ad,\phi}=0$ in the quasi-stationary state).
From Eq.~\eqref{eq:Lorentz force}, we see that the magnitude of the magnetic force can be estimated as $f_B \sim B^2/(4\pi \ell_B)$,
where $\ell_B$ is the length scale over which the magnetic field varies. Under these assumptions, we obtain the usual estimate for magnetic evolution under ambipolar diffusion as $t_{B} \sim t_{ad}$, where
\begin{equation}
t_{ad} \equiv \frac{R}{|\vec{v}_{ad\text{, Pol}}|} \sim \frac{4\pi \gamma_{cn}n_c n_n \ell_B R}{B^2} \label{eq:t_ad}\,.
\end{equation}
This derivation, however, has a few shortcomings. First, 
from Eqs.~\eqref{eq:force_balance_1} and \eqref{eq:v_ambipolar}, we see that a better estimate of the magnitude of the poloidal component of the ambipolar velocity is $v_{ad} \sim f_n/(\gamma_{cn}n_c n_n)$;
however, as demonstrated by \citet{Moraga2024_MagnetothermalEvolutionCores}, an arbitrary magnetic field can induce fluid forces larger than the Lorentz force, i.~e., $f_c \sim f_n \sim (\ell_c/\ell_B)f_{B}$, 
where we introduced a length scale
\begin{equation}
    \ell_c\equiv -\left[\frac{d}{dr}\ln(n_{c}/n_{n})\right]^{-1}\,, \label{eq:lc}
\end{equation}
which is a measure of the stable stratification due to the composition gradient in the NS core [more stratified for smaller $\ell_c$, see Fig.~\ref{fig:n-ratio}(c)]. Thus,
\begin{equation}
    v_{ad\text{, Pol}} \sim \frac{\ell_c}{\ell_B}\frac{f_B}{\gamma_{cn}n_c n_n} \label{eq:va estimate}\,.
\end{equation}
More importantly, the implicit assumption of fixed neutrons is not justified. As shown in \citet{Ofengeim2018_FastMagneticField}, and numerically checked by \citet{Castillo2020_TwofluidSimulationsMagnetic}, the neutron velocity can be much larger than the ambipolar diffusion velocity. Its magnitude can be estimated from Eqs.~\eqref{eq:neutron force} and \eqref{eq:vnr_anl} as
\begin{equation}
    v_{n\text{, Pol}}\sim \frac{\ell_c}{\ell_B} \frac{f_n}{ \gamma_{cn} n_c n_n} \sim \frac{\ell_{c}}{\ell_B}v_{ad\text{, Pol}}\,. \label{eq:vn-va}
\end{equation}
Typically, $\ell_{c}/\ell_{B}\gg 1$ (see Sect.~\ref{sec:results:comparison});
therefore, from Eqs.~\eqref{eq:va estimate} and \eqref{eq:vn-va}, we obtain a more accurate estimate of the long-term evolution timescale of the magnetic field under ambipolar diffusion
\begin{equation}
t_B \sim\frac{R}{v_{n\text{, Pol}}}\sim \left(\frac{\ell_{B}}{\ell_c}\right)^2\frac{4\pi \gamma_{cn}n_c n_n \ell_B R}{B^2} = \left(\frac{\ell_{B}}{\ell_c}\right)^2 t_{ad}\,. \label{eq:t_B}
\end{equation}

In the case of the FF approach, the FF force sets a shorter timescale in which the star relaxes to a state of hydromagnetic quasi-equilibrium. 
The velocity of the motion that causes such relaxation can be obtained from Eq.~\eqref{eq:force_balance_2}, which yields $f_{B}\sim\zeta n_n v_{n}$.
Thus, the configuration of the magnetic field and particles adjusts on a timescale
\begin{equation}
t_{\zeta B} \sim \frac{R}{v_{n}} \sim \frac{\zeta n_n R}{f_B} \sim \frac{4\pi\zeta n_n \ell_B R}{B^2}\,, \label{eq:t_zB}
\end{equation}
which should be identified with an Alfvén-like time and must be much shorter than the dynamical time $t_B$ of our interest for our simulations to accurately represent the long-term evolution.
Since, under axial symmetry, there is no toroidal gradient of the chemical potentials to balance the toroidal magnetic force, the magnetic field will rearrange itself on a timescale $t_{\zeta B}$ so the toroidal component of the force is null in the quasi-stationary state.

\subsection{Magnetic energy dissipation}
\label{sec:evolution:energy}

The loss of total magnetic energy ($U_B$) due to ambipolar diffusion can be computed by taking the derivative of the magnetic energy and then using Eq.~\eqref{eq:faraday}.
\begin{equation}
\begin{split}
	\dot U_B&=\frac{1}{4\pi}\int_{\mathcal{V}_\infty} \vec{B}\cdot\frac{\partial\vec{B}}{\partial t}\,d\mathcal{V}  \\
	&=\frac{1}{4\pi}\int_{\mathcal{V}_\text{core}} \vec{B}\cdot \vec{\nabla}\times\left(\vec{v}_c\times\vec{B}\right)\,d\mathcal{V} + \dot U_{B\text{, Ext}}\,,
\end{split}
\end{equation}
where $\mathcal{V}_\infty$ stands for all space, and $U_{B\text{, Ext}}$ is the magnetic energy outside of the core.
Integrating this expression by parts and using Eq.~\eqref{eq:Lorentz force}, we obtain
\begin{equation}
\dot U_B=-\int_{\mathcal{V}_\text{core}}\vec{v}_c\cdot\vec{f}_{B}\,d\mathcal{V}\,,
\end{equation}
which is the net power transfer between the fluids and the magnetic field.
Note that the boundary term from the integration by parts (i.~e., minus the Poynting flux integrated over the surface of the core) cancels with $\dot U_{B\text{, Ext}}$.

In the case of the FF approach, the latter can be further disassembled using Eqs.~\eqref{eq:v_ambipolar} and \eqref{eq:v_neutrons}
\begin{equation}
\begin{split}
\dot U_B =& -\int_{\mathcal{V}_\text{core}}\vec{v}_{ad}\cdot\left(\gamma_{cn}n_c n_n\vec{v}_{ad} -\vec{f}_{c}\right)\,d\mathcal{V} \\
&-\int_{\mathcal{V}_\text{core}}\vec{v}_n\cdot\left(\zeta n_n\vec{v}_n-\vec{f}_{n}-\vec{f}_{c}\right)\,d\mathcal{V}\,,
\end{split}
\end{equation}
Here, we identify the energy loss due to binary collisions between charged particles and neutrons as
\begin{equation}
L_{ad}=\int_{\mathcal{V}_\text{core}} \gamma_{cn}n_c n_n|\vec{v}_{ad}|^2 \,d\mathcal{V} \,, \label{eq:L_ad}\\
\end{equation}
and the energy loss due to the FF on the neutrons
\begin{equation}
L_{\zeta}=\int_{\mathcal{V}_\text{core}} \zeta n_n|\vec{v}_n|^2 \,d\mathcal{V} \,. \label{eq:L_z}
\end{equation}
The rate at which energy goes into (or comes from) the charged particles can be identified as
\begin{equation}
\dot U_c =-\int_{\mathcal{V}_\text{core}} \vec{v}_c\cdot\vec{f}_{c} \,d\mathcal{V}  \label{eq:dU_c}\,,
\end{equation}
This expression can be re-arranged by replacing Eq.~\eqref{eq:charged force} and integrating by parts. Then, using Eq.~\eqref{eq:continuity_c2} yields
\begin{equation}\label{eq:dU_c2}
\begin{split}
\dot U_c &=\int_{\mathcal{V}_\text{core}} n_c\chi_c\vec{\nabla}\mu\cdot\vec{v}_c \,d\mathcal{V}  \\
&=-\int_{\mathcal{V}_\text{core}} n_c\frac{\delta\mu_c}{c^2}\vec{\nabla}\Phi\cdot\vec{v}_c \,d\mathcal{V} \,.
\end{split}
\end{equation}
In the last equality, we have used Eq.~\eqref{eq:hydrostatic equilibrium}.
Thus, $-\dot U_c$ could be interpreted as the mechanical work per unit time done by gravity on the charged particle fluid.
Analogously, the work per unit time done by gravity on the neutrons can be identified as
\begin{equation}
-\dot U_n =\int_{\mathcal{V}_\text{core}} n_n\frac{\delta\mu_n}{c^2}\vec{\nabla}\Phi\cdot\vec{v}_n \,d\mathcal{V} 
\label{eq:dU_n}\,.
\end{equation}
$\dot U_c$ and $\dot U_n$ can be positive or negative.
Note, however, that these terms are a direct consequence of the adopted Newtonian approximation 
(in particular, the way we account for redshifts) and they do not appear in the fully relativistic treatment (\citealt{Gusakov2017_EvolutionMagneticField}; Moraga et al., in preparation). 

Combining the equations above, the net rate of change of the total magnetic energy can be written as
\begin{equation}
    \dot U_B = -L_{ad}-L_{\zeta}-\dot U_c-\dot U_n\,. \label{eq:dU_B}
\end{equation}

\section{Results}
\label{sec:results}

\subsection{Comparison between the different approaches}
\label{sec:results:comparison}

In this section, we compare the different methods described earlier by evaluating the neutron and ambipolar velocities.
The different methods are compared to the semi-analytical velocities obtained for the magnetic field models studied in this work.

In Fig.~\ref{fig:models}, we show the relative chemical potential perturbations, $\chi_n$ and $\chi_c$, and the velocity fields for each magnetic field model, evaluated with the explicit semi-analytical approach described in Sect.~\ref{sec:semi-anallytical:explicit} using numerical data from the HHJ EoS. 
Note, however, that model IV is not in quasistationary equilibrium since, for this model, $f_{B,\phi}$ is non-zero. Thus, strictly speaking, the procedure outlined in Sect.~\ref{sec:semi-anallytical:explicit} does not apply.
Aside from this caveat, we see that the neutron velocity is much larger than the ambipolar velocity for all magnetic field models studied and that this difference becomes larger as the magnetic field has a more complex structure, which is consistent with Eq.~\eqref{eq:vn-va}.
This can be seen more clearly in Table~\ref{tab:rms vn/va}, where the non-purely dipolar and non-equatorially symmetric
models III and IV yield the largest ratios $\langle\vec{v}_n\rangle/\langle\vec{v}_{ad}\rangle$.

\begin{table}
	\caption{Rms values of the velocities displayed in the last two rows of Fig.~\ref{fig:models} expressed in code units (see Table~\ref{tab:normalization}).}
	\label{tab:rms vn/va}
	\centering
	\begin{tabular}{lccc}
		\hline\hline
		 & $\langle\vec{v}_n\rangle$ & $\langle\vec{v}_{ad}\rangle$ & $\langle\vec{v}_n\rangle/\langle\vec{v}_{ad}\rangle$\\
		\hline
		Model I & 11.2 & 0.452 & 24.8\\
		Model II & 51.0 & 2.45 & 20.8\\
		Model III & 661 & 6.40 & 103\\
        Model IV & 118 & 2.29 & 51.3\\
		\hline
	\end{tabular}
\end{table}

\subsubsection{Solutions for the pseudo-spectral approach}
\label{sec:results:comparison:spectral}

As described in Sect.~\ref{sec:semi-anallytical:spectral}, the system of equations from Sect.~\ref{sec:model} can be rewritten for a direct solution with a pseudo-spectral code (using Eqs.~\ref{eq:v_ambipolar23}, \ref{eq:chinl_geq1}, \ref{eq:chicl_geq1}, \ref{eq:v_n23}, and \ref{eq:Psi_l}). The solution is implemented with \texttt{Dedalus}.

\begin{table}
	\caption{Coefficients for the fit of $x=n_n,n_c,\mu$, and $\gamma_{cn}$, using tabulated data from the HHJ EoS, expressed in code units (see Table~\ref{tab:normalization}). Each radial function is expressed as $x_\text{fit}(r)=(\sum_{i=0}^3 a_{2i}r^{2i})/(\sum_{i=0}^3 b_{2i}r^{2i})$.}
	\label{tab:fit}
	\centering
	\begin{tabular}{lcccc}
		\hline\hline
		 & $n_n$ & $n_c$ & $\mu$ & $\gamma_{cn}$ \\
		\hline
		$a_0$ & 10.958 & 1 & 1 & 1 \\
		$a_2$ & -24.880 & -1.5433 & -1.1982 & -1.5004 \\
		$a_4$ & 18.206 & 0.39369 & 0.33677 & 0.30926\\
		$a_6$ & -4.2368 & 0.16089 & -0.056229 & 0.20444\\
		$b_0$ & 1 & 1 & 1 & 1 \\
		$b_2$ & -1.5185 & -0.36226 & -0.94247  & -2.5542\\
		$b_4$ & 0.46929 & -0.33698 & 0.027858 & 2.1541\\
		$b_6$ & 0.074542  & -0.098699 &  0.015922 & -0.59869\\
		\hline
	\end{tabular}
\end{table}

\begin{figure}
	\centering
	\includegraphics[width=\linewidth]{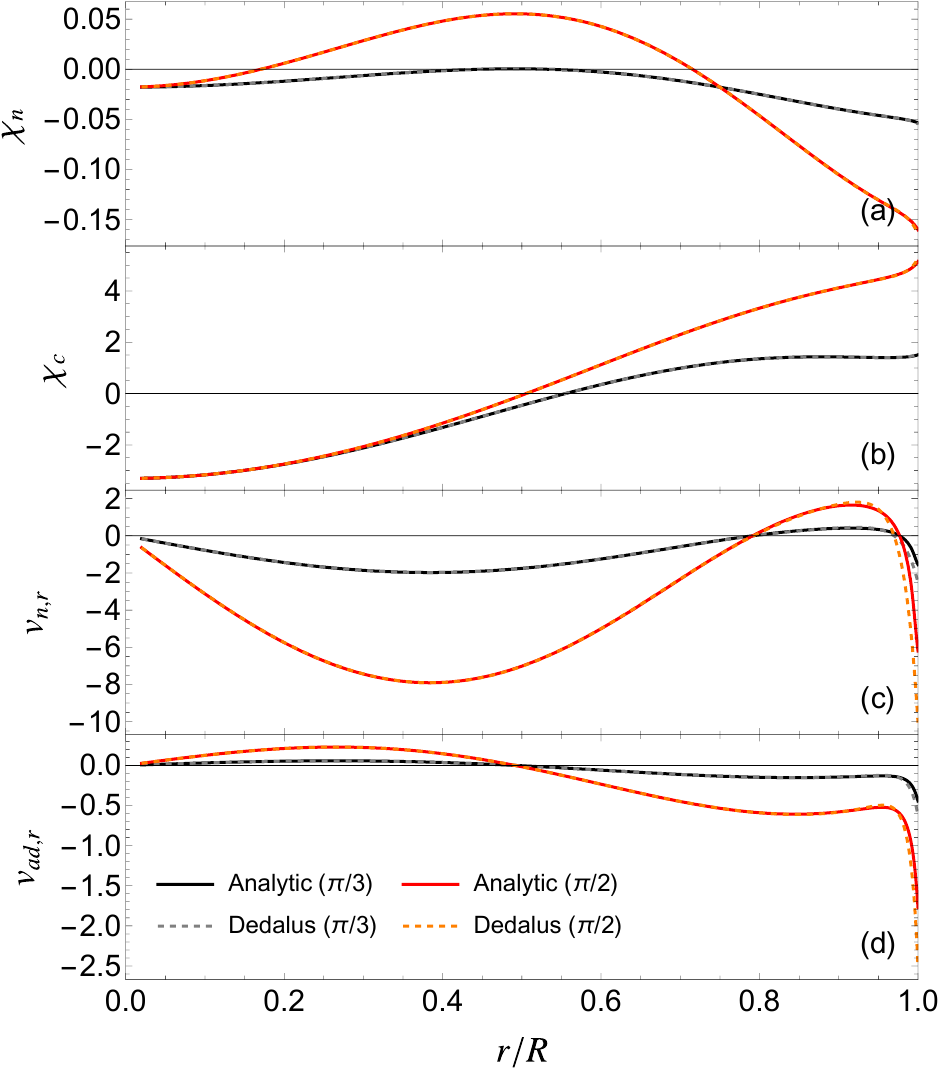}
	\caption{Comparison of the pseudo-spectral approach (dashed line) and the semi-analytical solution obtained using the explicit approach (continuous line) for magnetic field model I. The solution from the pseudo-spectral approach uses a polynomial fit for the relevant background radial profiles.
    From top to bottom: (a) $\chi_n$, (b) $\chi_c$, (c) $v_{n,r}$, and (d) $v_{ad,r}$. All quantities are in code units, plotted as functions of $r$ for $\theta=\pi/3$, and $\theta=\pi/2$.}
	\label{fig:comp-dedalus-model_i}
\end{figure}

\begin{figure}
	\centering
	\includegraphics[width=\linewidth]{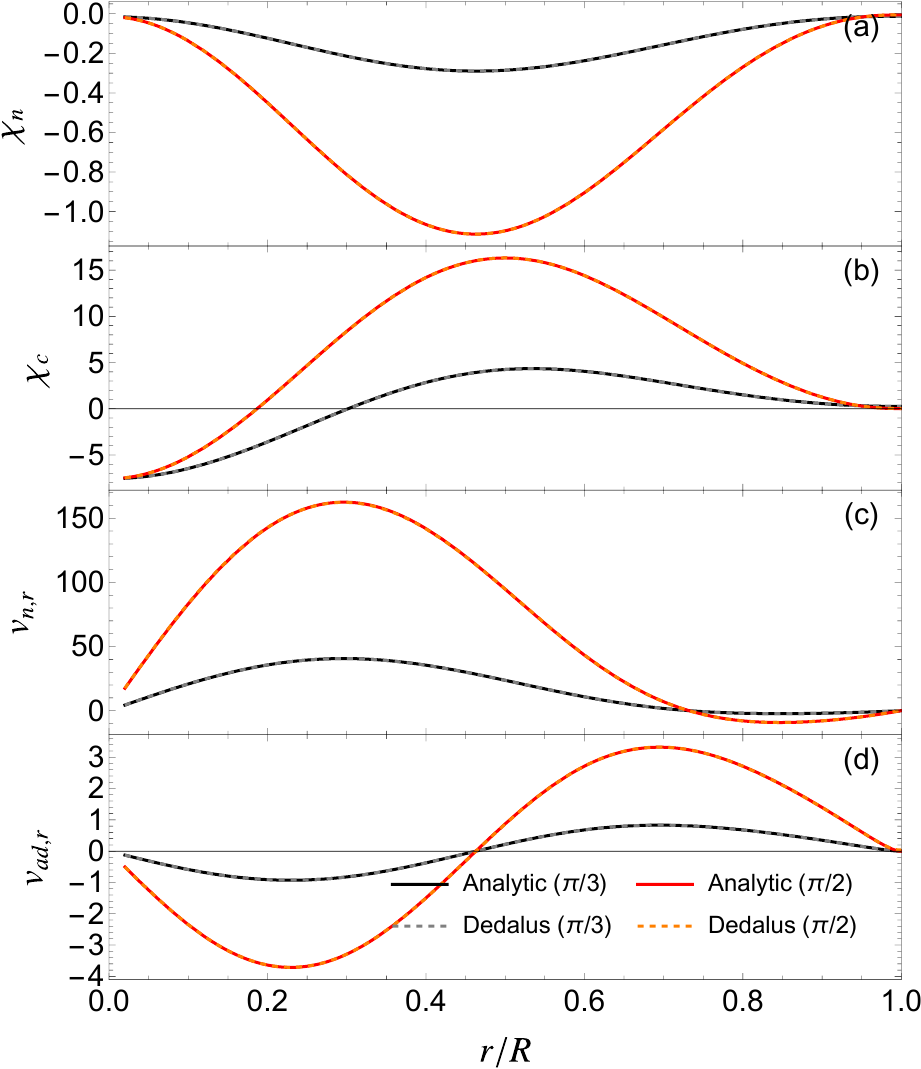}
	\caption{Comparison of the output of the pseudo-spectral approach (dashed line) and the semi-analytical solution obtained using the explicit approach (continuous line) for magnetic field model II. The solution from the pseudo-spectral approach uses a polynomial fit for the relevant background radial profiles.
    From top to bottom: (a) $\chi_n$, (b) $\chi_c$, (c) $v_{n,r}$, and (d) $v_{ad,r}$. All quantities are in code units, plotted as functions of $r$ for $\theta=\pi/3$, and $\theta=\pi/2$.}
	\label{fig:comp-dedalus-model_ii}
\end{figure}

\begin{figure}
	\centering
	\includegraphics[width=\linewidth]{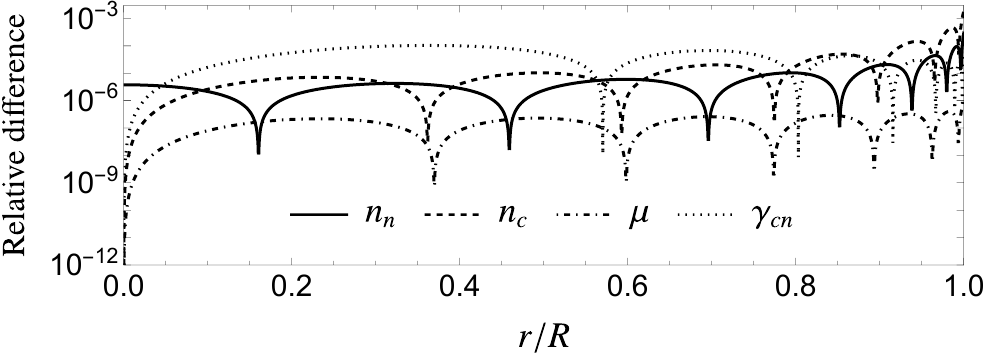}
	\caption{
    Relative difference between the analytical fit of the radial profiles obtained for the HHJ EoS (see Table~\ref{tab:fit}) and the data extracted from tables. i.~e., $|x_\text{fit}(r)/x(r)-1|$ for $x=n_n,n_c,\mu$, and $\gamma_{cn}$.
    }
	\label{fig:comp. HHJ fit}
\end{figure}

Figs.~\ref{fig:comp-dedalus-model_i} and \ref{fig:comp-dedalus-model_ii} show radial profiles at two angles for the chemical potential perturbations and for the neutron and ambipolar velocities using magnetic field models I and II. 
For the pseudo-spectral approach, analytical fitting formulae for the background quantities ($n_n$, $n_c$, $\gamma_{cn}$, and $\mu$; hereafter, HHJ-fit) were used (see Table~\ref{tab:fit}), since using numerical data from the HHJ EoS yields incorrect results for the velocities at the crust-core interface.
The small difference between the HHJ-fit model and HHJ is shown in Fig.~\ref{fig:comp. HHJ fit}.
In Figs.~\ref{fig:comp-dedalus-model_i} and \ref{fig:comp-dedalus-model_ii} we see
there is very good agreement between the output of \texttt{Dedalus} and the semi-analytical solution evaluated using the explicit approach in most of the star.
However, there are small discrepancies close to the crust-core interface.
This is consistent with the fact that $n_c$, while overall quite similar in the HHJ-fit and HHJ models, 
differs up to $\sim 0.2\%$ close to the crust-core interface (see Fig.~\ref{fig:comp. HHJ fit}).

\subsubsection{Solutions for the FF approach}
\label{sec:results:comparison:FF}

As demonstrated in Sect.~\ref{sec:fictitious_friction}, the velocities obtained in the FF approach will differ from the real ones, with their difference increasing with $\zeta$.
From the expression used in this approach for $\vec{v}_n$ (Eq.~\ref{eq:v_neutrons}), it is not evident for what value of $\zeta$ the obtained velocities (and other relevant variables) will resemble those obtained by the explicit approach with reasonable accuracy. 
It is also not obvious whether this value would be large enough so we can numerically invert the matrix $\Lambda$, which, as mentioned in Sect.~\ref{sec:numerical:FD}, becomes singular for $\zeta=0$.
Therefore, a numerical test for different values of $\zeta$ becomes necessary. 

\begin{figure}
	\centering
	\includegraphics[width=\linewidth]{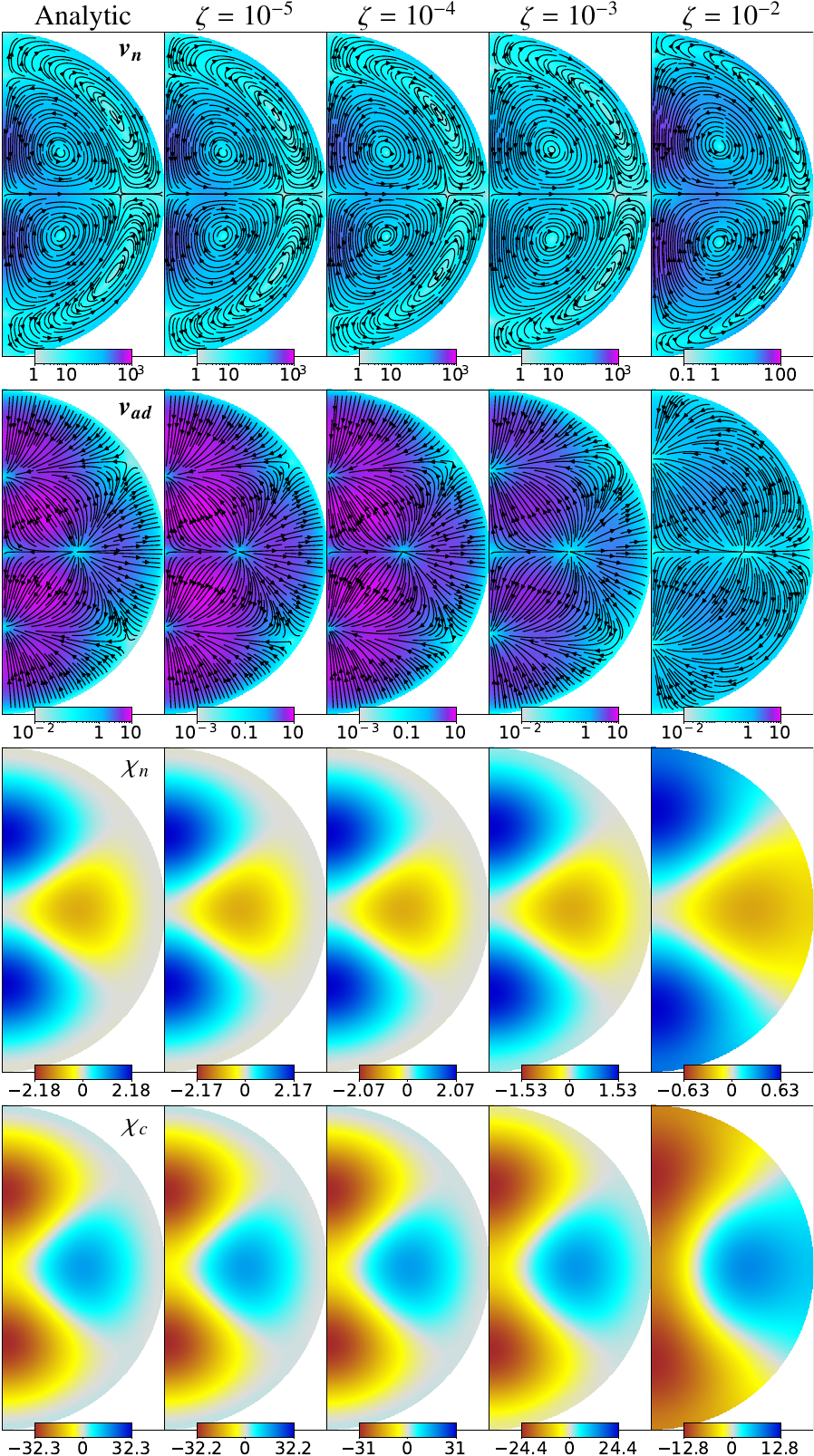}
	\caption{Comparison between the explicit semi-analytical evaluation of the velocities and relative chemical potential perturbations (corresponding to $\zeta=0$), and the numerical calculation of the same quantities using the FF approach with different values of $\zeta$. For the velocities $\vec{v}_{n}$ and $\vec{v}_{ad}$, lines represent the direction of their poloidal component and color its magnitude. This comparison is made for magnetic field model II, using the HHJ EoS. The computation is done over an equally spaced grid ($u=1$ in equation \ref{eq:r_i}) of resolution $N_r=60$, $N_\theta=91$, using an analytical expression for $\vec{f}_B$ to minimize numerical errors.}
	\label{fig:z-comp model II}
\end{figure}

In Fig.~\ref{fig:z-comp model II}, we show a comparison between four very different values of $\zeta$ and the semi-analytical result obtained using the explicit approach, which corresponds to $\zeta=0$ (hereafter, we refer to $\zeta$ in the dimensionless units of Table~\ref{tab:normalization}). 
The computation is done using magnetic field model II, as this is specifically built so that $v_{n,r}=v_{ad,r}=0$ at the crust-core interface, which in the FF approach is taken as a boundary condition. It can be observed that even for values as high as $\zeta=10^{-2}$ the velocities and relative chemical potential perturbations are qualitatively similar to their semi-analytical counterpart in most of the volume, but they differ in magnitude.
For $\zeta\leq 10^{-3}$, the different variables also show similar qualitative behavior in the whole volume, but the difference in their magnitude decreases. For instance, for $\zeta=10^{-3}$, the maximum neutron velocity is $\sim 47\%$ smaller than the semi-analytical value, suggesting that this value of $\zeta$ may still be too large for an accurate evaluation, for $\zeta=10^{-4}$ the maximum velocity is $\sim 10\%$ smaller, while for $\zeta=10^{-5}$, it is just $\sim 1.2\%$ smaller.
In practice, as discussed in Sect.~\ref{sec:results:evolution:FF_convergence}, for time-evolution simulations, we are using values of $\zeta$ in the range $10^{-4}$--$10^{-3}$, as smaller values yield a much longer integration time.

\begin{figure}
	\centering
	\includegraphics[width=\linewidth]{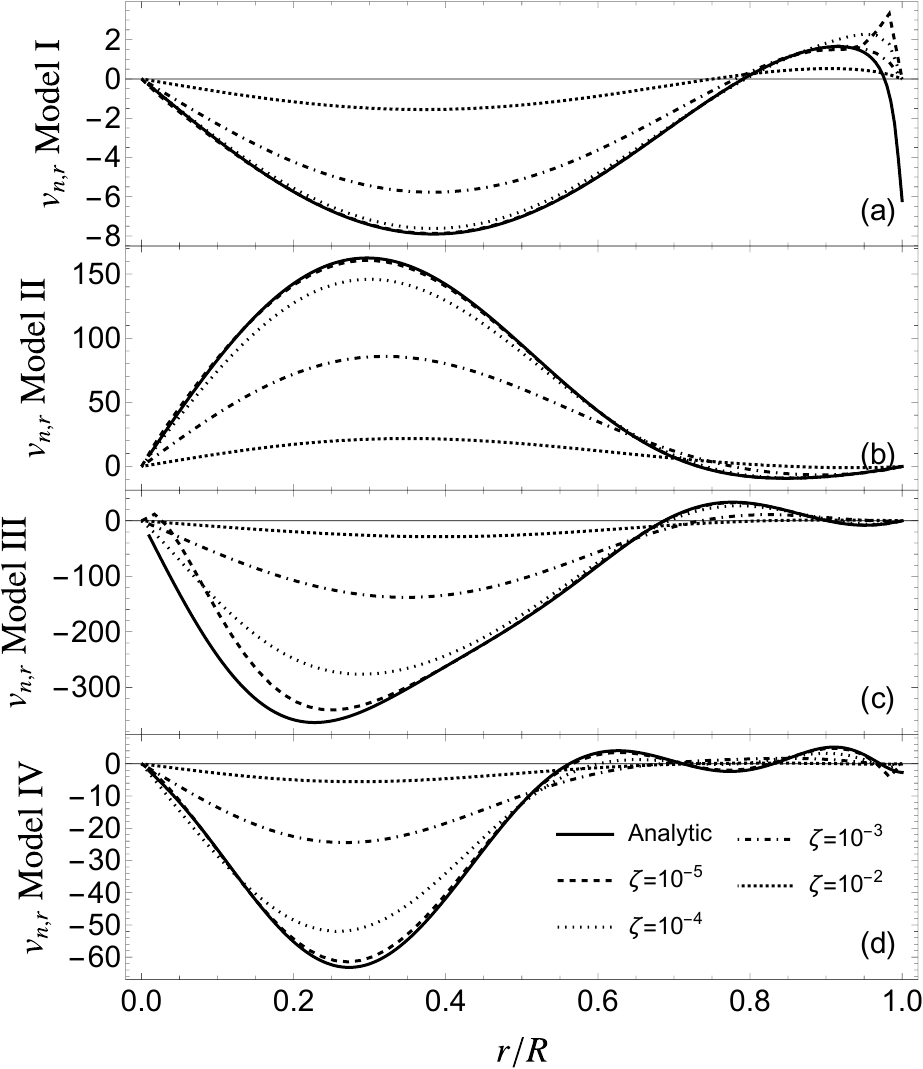}
	\caption{Radial component of the neutron velocity, $v_{n,r}$, for $\theta=\pi/2$ as a function of $r$ for magnetic field models I, II, III, and IV, all with the HHJ EoS. The evaluation is done with the explicit approach for $\zeta=0$ (no FF) and numerically using the FF approach with different values of $\zeta>0$.}
	\label{fig:z-comp models radial}
\end{figure}

The same behavior can also be observed in Fig.~\ref{fig:z-comp models radial}, where we plot $v_{n,r}$ as a function of $r$ for $\theta=\pi/2$ for different magnetic field models and for different values of $\zeta$, finding convergence for small $\zeta$ for all models. However, for model I, the radial velocity near the crust-core interface obtained with the FF approach differs from the one obtained using the explicit approach because the latter is not null at the crust-core interface, whereas null radial velocities are enforced as boundary conditions in the FF approach. Note, however, that the velocities are substantially different only in a thin layer of thickness $\sim 0.05\,R$.
Note that for model III the curve using $\zeta=10^{-5}$ has a sudden change of sign close to $r=0$, which is not observed for the other models, whose curves go smoothly to zero.
This suggests that such a value of $\zeta$ may be too small, as it yields an almost-singular $\Lambda$ matrix (see Eq.~\ref{eq:Lambda LS}).

\begin{figure}
	\centering
	\includegraphics[width=\linewidth]{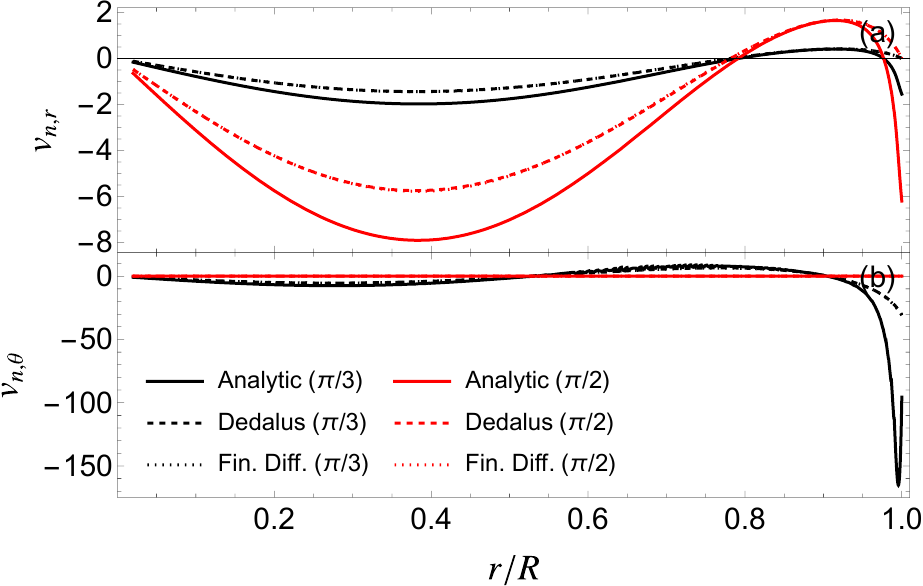}
	\caption{(a) Radial and (b) polar components of the neutron velocity for $\theta=\pi/3$ and $\theta=\pi/2$ as functions of $r$, for magnetic field model I using the HHJ EoS. The evaluation is done semi-analytically (corresponding to $\zeta=0$; continuous line) and numerically, using the FF approach with \texttt{Dedalus} (dashed line) and our finite difference code (dotted line), both using $\zeta=10^{-3}$.}
	\label{fig:Dedalus_model_i_z0.001_HHJ-vn}
\end{figure}

\begin{figure}
	\centering
	\includegraphics[width=\linewidth]{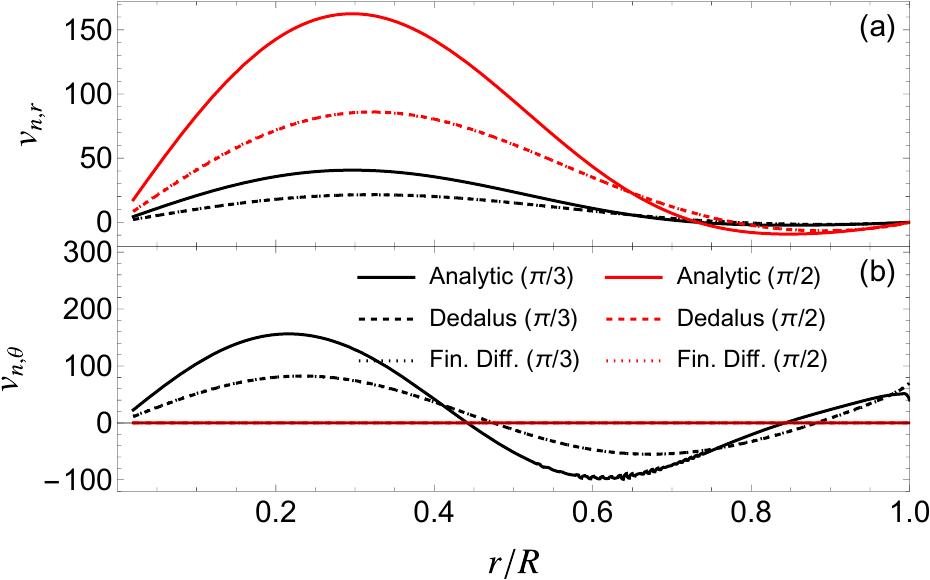}
	\caption{(a) Radial and (b) polar components of the neutron velocity for $\theta=\pi/3$ and $\theta=\pi/2$ as functions of $r$ for magnetic field model II using the HHJ EoS. The evaluation is done semi-analytically (corresponding to $\zeta=0$; continuous line) and numerically, using the FF approach with \texttt{Dedalus} (dashed line) and our finite difference code (dotted line), both using $\zeta=10^{-3}$.}
	\label{fig:Dedalus_model_ii_z0.001_HHJ-vn}
\end{figure}

The equations for the FF approach can also be solved within the pseudo-spectral code \texttt{Dedalus}.
Figs.~\ref{fig:Dedalus_model_i_z0.001_HHJ-vn} and~\ref{fig:Dedalus_model_ii_z0.001_HHJ-vn} compare the neutron velocity field profiles obtained through the explicit approach and the artificial friction approach, where the latter was implemented using both a finite difference method and \texttt{Dedalus}.
The figures show magnetic field models I and II at a given angle and with the value of $\zeta=10^{-3}$, using the HHJ EoS. 
Unlike the pseudo-spectral approach, no explicit analytical fit for the relevant background radial profile is used.
The results show a very good agreement between the pseudo-spectral and the finite difference implementation of the FF approach.
However, for time-evolving simulations, we restrict to the finite difference implementation, as our code was developed exactly for this purpose and has already been extensively tested.

\subsection{Hydro-magnetic evolution}
\label{sec:results:evolution}

In this work, we have explored different methods of obtaining the neutron and ambipolar velocities for a given magnetic field configuration, with the intention of finding a reliable method to evolve the magnetic field in time. Sadly, while we can solve the equations from Sect.~\ref{sec:model} for analytically given magnetic field configurations using the semi-analytical approaches of Sect.~\ref{sec:semi-anallytical}, we have not been successful in our efforts to iterate such procedures in time to evolve the magnetic field.
The reason for this seems to be related to two factors: First, the evaluation of the neutron velocity involves a large number of spatial derivatives of the magnetic field (five derivatives for $v_{n,\theta}$; see Sect.~\ref{sec:semi-anallytical:explicit}), which require high resolution to be solved properly. For a fixed magnetic field configuration, these can be computed to a reasonable accuracy, but, in time-evolving simulations, numerical errors in their calculation accumulate.
Second, for an arbitrary magnetic field, we do not have the freedom to fix boundary conditions for the velocities, whose values at the crust-core interface are completely determined by the magnetic field. Furthermore, even if the initial seed for the magnetic field is chosen to satisfy a given condition (for instance, null radial velocities), it is unclear if this magnetic field will evolve in time in such a way that the boundary conditions continue to be satisfied. 

On the other hand, the FF approach offers a reliable way of evolving the magnetic field in time. Also, in the present version of the code, we have resolved some of the main caveats in the work of \citet{Castillo2020_TwofluidSimulationsMagnetic}. Namely, in that version, time derivatives in the continuity Eqs.~\eqref{eq:continuity_n} and \eqref{eq:continuity_c} were not neglected and were used to evolve the density perturbations $\delta n_n$ and $\delta n_c$ in time, which requires resolving an extremely short timescale $t_{\zeta p} \sim \chi_0\, t_{\zeta B} \sim 10^{-10}(B/10^{13}G)^2\,t_{\zeta B}$, which is the analog of the propagation time of sound waves (p-modes) when inertial effects are taken into account. Thus, performing simulations in which we could numerically resolve both timescales, with the available computational resources, required to artificially increase the magnetic field strength up to unrealistic values $B\gtrsim 10^{17}G$, so $t_{\zeta p}$ and $t_{\zeta B}$ could get closer.
In the current version, however, time derivatives in the continuity equations are neglected, in practice replacing the need to follow the evolution of density perturbations in favor of directly solving for the chemical potential perturbations from the linear system of Eqs.~\eqref{eq:chi solve}.
Therefore, the shortest timescale to be resolved in the new scheme is $t_{\zeta B}$, and the two relevant timescales, $t_{\zeta B}$ and $t_{ad}$, scale $\propto B^{-2}$, allowing us to perform simulations for one field strength and scale them to any other, as done by \citet{Moraga2024_MagnetothermalEvolutionCores} for the magneto-thermal evolution in the high-temperature strong-coupling regime.
Thus, in this subsection, we will restrict to time-evolving simulations using the FF approach.

\subsubsection{Convergence for different values of \texorpdfstring{$\zeta$}{ζ}}
\label{sec:results:evolution:FF_convergence}

The value of $\zeta$ used does not only determine how close we are to the real velocities. In a time-evolving simulation, it also fixes the ratio between the dynamical timescales $t_{ad}$ and $t_{\zeta B}$.
For a star of $B\sim 10 ^{13}\,\text{G}$ we have an Alfvén time of $t_{A}\sim 10\,\text{s}$.
Since we have identified $t_{\zeta B}$ with an Alfvén-like timescale, a realistic value of $\zeta$ that satisfies $t_{A}\sim t_{\zeta B}$ would be $\zeta=t_{\zeta B}/t_{ad} \sim 10^{-16}$ (in dimensionless units), making it impossible to resolve both timescales in a simulation. 
Therefore, in our simulations, we use much larger values of $\zeta$ so that both dynamical timescales are close enough to be resolved.

\begin{figure}
	\centering
	\includegraphics[width=\linewidth]{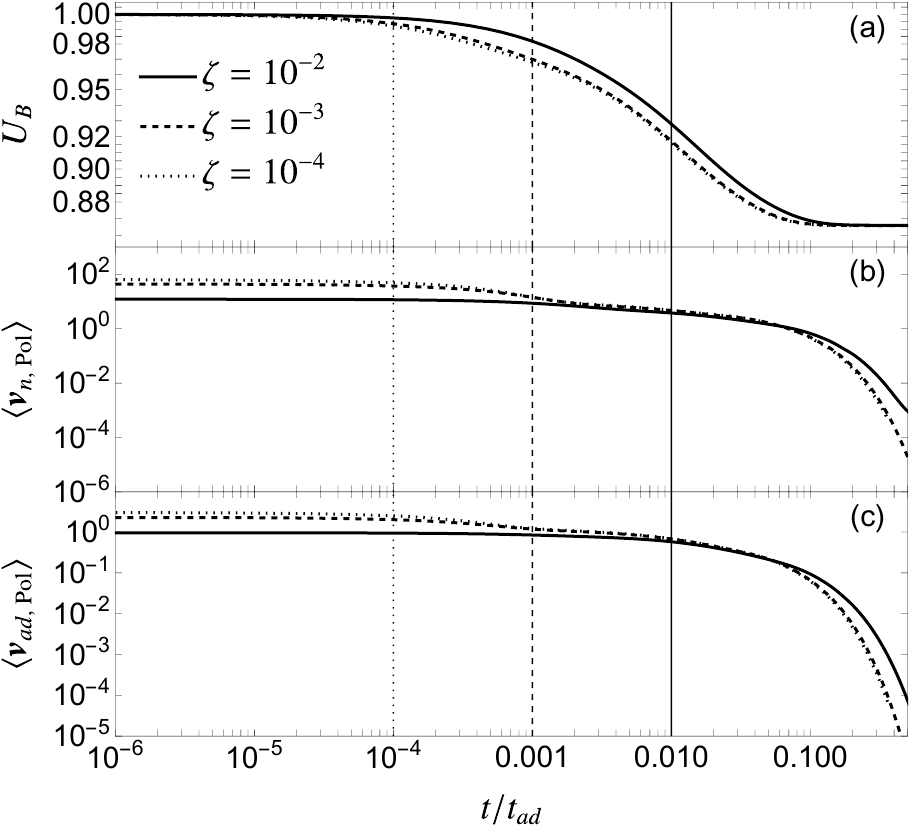}
	\caption{Time-evolution for magnetic field model II with background NS provided by the EOS HHJ, using different values of $\zeta$. (a)~Evolution of the total magnetic energy, normalized by its initial value. (b) and (c)~evolution of the rms values of the neutron velocity and ambipolar velocity, respectively. Vertical lines mark the value of $t_{\zeta B}/t_{ad}$ for each simulation.}
	\label{fig:z comp model II}
\end{figure}

To determine the most appropriate range for $\zeta$, we have performed simulations for different values, using magnetic field model II as the initial condition. 
These simulations were done over a non-homogeneous grid of resolution $N_r=60$, $N_\theta=91$, setting $u=2/3$ in Eq.~\eqref{eq:r_i}.
This is done so the grid cells become larger close to the center, yielding a less restrictive Courant–Friedrichs–Lewy condition \citep{Courant1928_UberPartiellenDifferenzengleichungen}, at the expense of losing numerical accuracy close to the center of the star.
To facilitate comparison with the results of \citet{Castillo2020_TwofluidSimulationsMagnetic}, hereafter we computed $t_{ad}$ and $t_{\zeta B}$ (Eqs.~\ref{eq:t_ad} and \ref{eq:t_zB}) using $B=\langle \vec{B}(t=0)\rangle$, $\ell_B=R/16$, and stellar parameters ($n_n$, $n_c$, $\gamma_{cn}$) evaluated at the center of the star (see Table~\ref{tab:stellar parameters}).
The results are in Fig.~\ref{fig:z comp model II}. It can be seen in the figure that the simulations show very similar behavior. Also, as the value of $\zeta$ gets smaller, simulations appear to converge to the same curve, particularly for $\zeta\lesssim 10^{-3}$, which suggests that this value of $\zeta$ is good enough for our purposes.

\begin{figure}
	\centering
	\includegraphics[width=\linewidth]{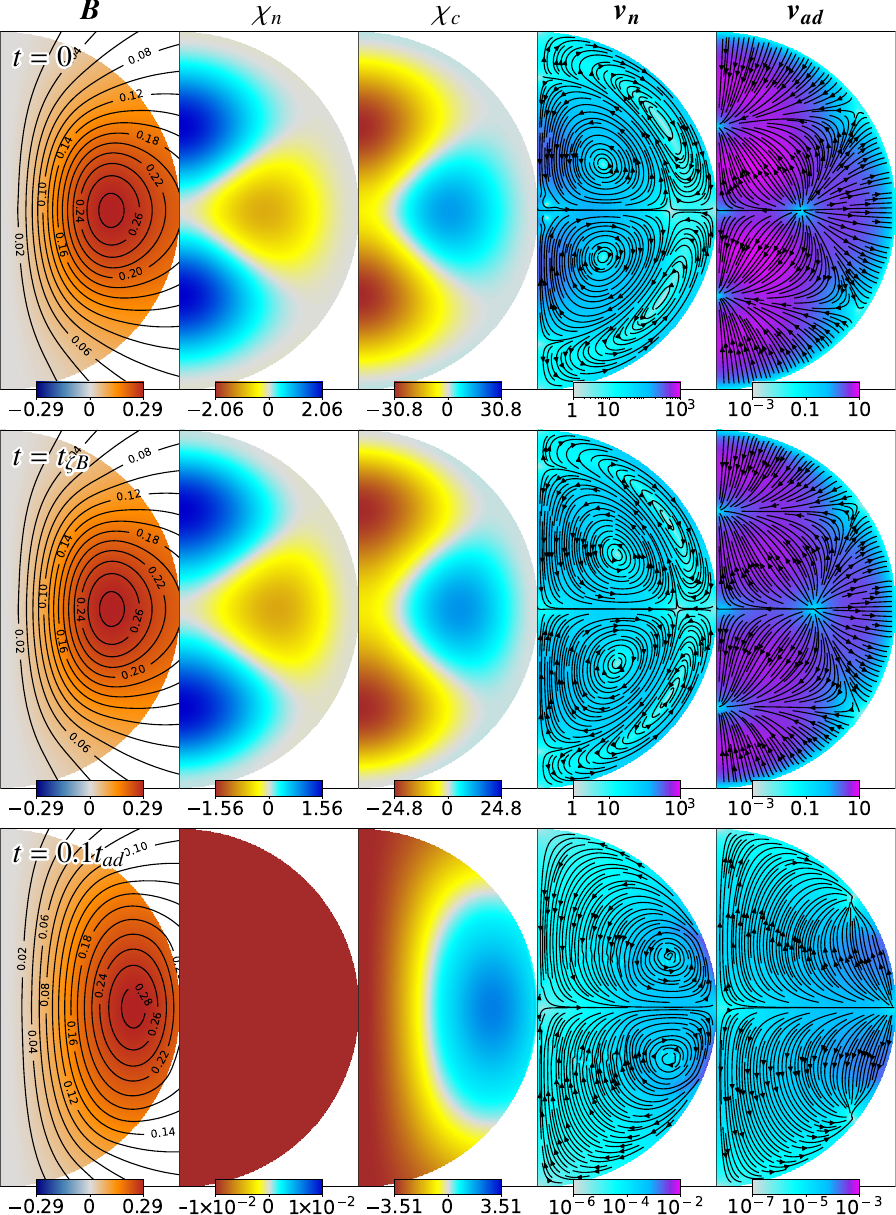}
	\caption{
	Evolution of magnetic field model II, using $\zeta=10^{-4}$. We used a grid of $N_r=60$ radial steps and $N_\theta=91$ polar steps inside the core, $N_{\text{Exp}}=40$ external multipoles, $u=2/3$, and the HHJ EoS. From left to right: Configuration of the magnetic field, where lines represent the poloidal magnetic field (labeled by the magnitude of $\alpha$)
    and colors the poloidal potential $\alpha$; $\chi_n$ and $\chi_c$, both in units of $\chi_0=B_0^2/4\pi n_0\mu_0$; 
    neutron velocity, $\vec{v}_n$, and ambipolar diffusion velocity, $\vec{v}_{ad}$, where arrows represent the direction and colors the magnitude normalized to $R/t_0$. Rows correspond to different times: $t=0$, $t=t_{\zeta B}=10^{-4}\,t_{ad}$, and $t=0.1\,t_{ad}$.
	}
	\label{fig:model II z1e-4 evolution}
\end{figure}

\begin{figure}
	\centering
	\includegraphics[width=\linewidth]{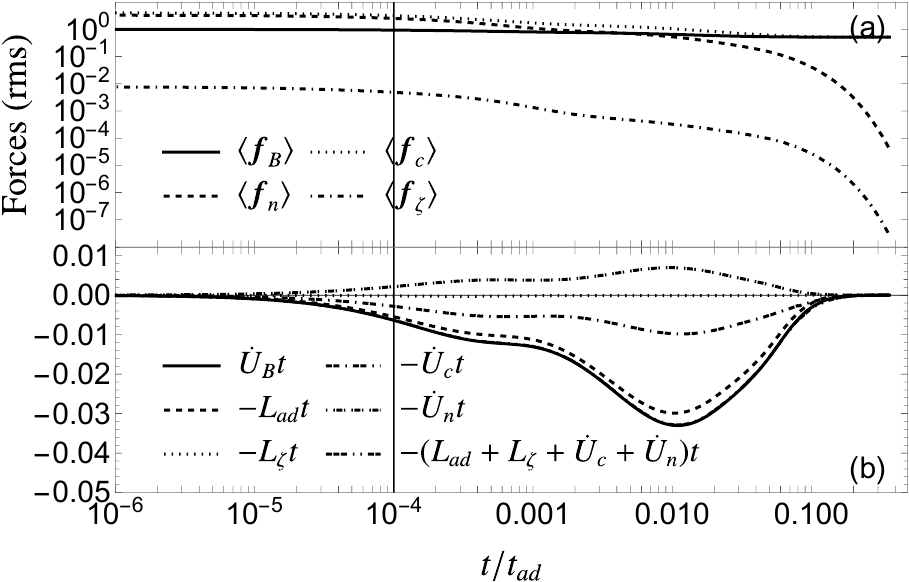}
	\caption{For the simulation of Fig.~\ref{fig:model II z1e-4 evolution}: (a)~Time evolution of the rms force densities $\langle \vec{f}_{B} \rangle$, $\langle \vec{f}_{n} \rangle$, $\langle \vec{f}_{c} \rangle$, and $\langle \vec{f}_{\zeta} \rangle$, all normalized by $\langle \vec{f}_{B}(t=0)\rangle$.
    (b) Energy conservation. All terms are in units of the total initial magnetic energy.
    The vertical line marks $t_{\zeta B}/t_{ad}$.
	}
	\label{fig:model II z1e-4 forces}
\end{figure}

In Sect.~\ref{sec:results:comparison:FF}, we showed how, for a fixed magnetic field, the FF approach yields velocity fields close to the semi-analytical ones only when $\zeta \lesssim 10^{-5}$, which seems to contradict our previous statement. However, there is a key difference.
In our tests with a fixed magnetic field, the only mechanism to diminish the FF force compared to the real physical forces is to take a smaller value of $\zeta$.
In time-evolving simulations, however, the magnetic field will adjust on a timescale $t_{\zeta B}$ to a configuration in which the FF force progressively decreases, thereby yielding velocities that more closely approximate the real velocities.
To show this more clearly, we focus on the simulation with $\zeta=10^{-4}$. Snapshots of the evolution can be seen in Fig.~\ref{fig:model II z1e-4 evolution}.
In Fig.~\ref{fig:model II z1e-4 forces}(a), we observe that the FF force is always weak and becomes progressively weaker compared to the other forces, except for $\vec{f}_{n}$, which also decreases over time due to ambipolar diffusion.
Fig.~\ref{fig:model II z1e-4 forces}(b) shows very good agreement between the left and right-hand sides of Eq.~\eqref{eq:dU_B}, where the left-hand side is computed explicitly as the numerical time derivative of the magnetic energy.

\begin{figure}
	\centering
	\includegraphics[width=\linewidth]{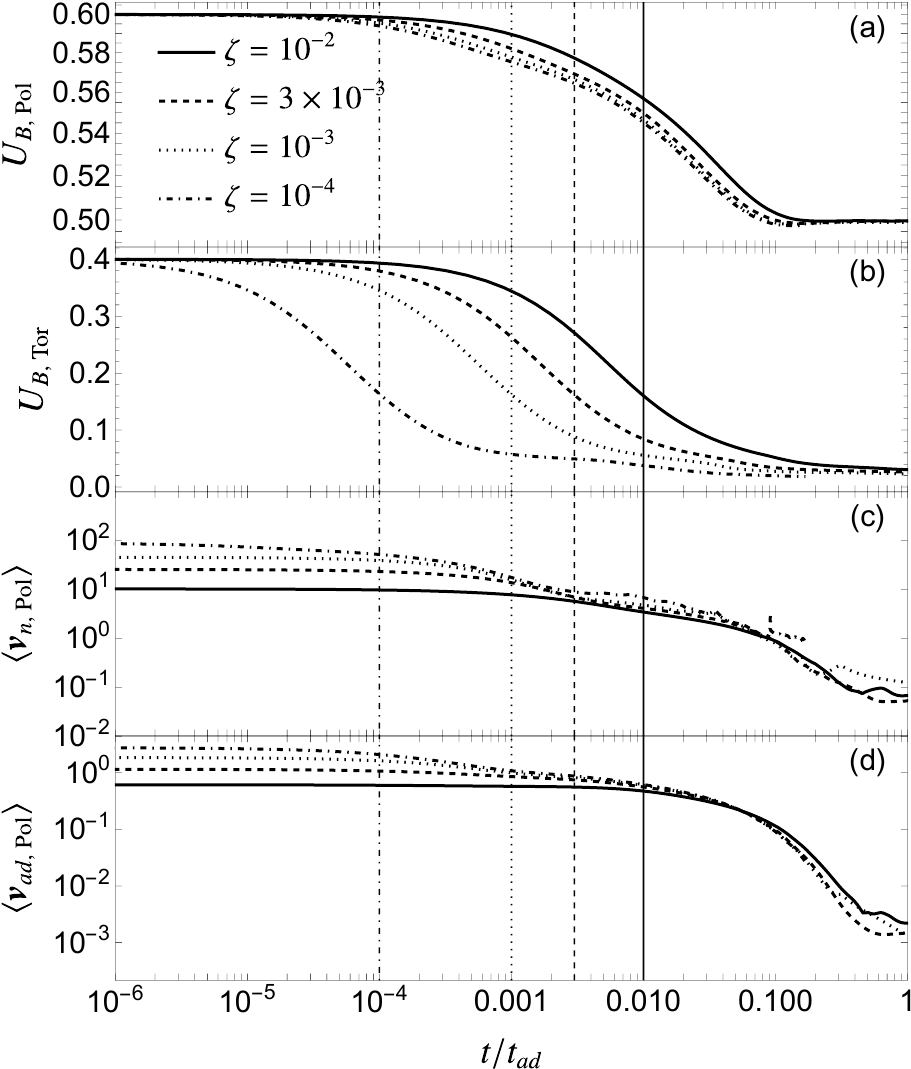}
	\caption{Plots for magnetic field model IV using different values of $\zeta$: (a) and (b)~Evolution of the poloidal and toroidal magnetic energy, respectively, normalized by the initial value of the magnetic energy of the star. (c) and (d)~evolution of the rms poloidal neutron and ambipolar velocity. Vertical lines correspond to the value of $t_{\zeta B}/t_{ad}$ in each simulation.}
	\label{fig:z comp model IV}
\end{figure}

The results described in this section apply not only to equatorially-symmetric magnetic field configurations. Figure~\ref{fig:z comp model IV} shows the evolution of the magnetic energy and rms velocities for different values of $\zeta$ using the magnetic field model IV, which is more complex, as it is not equatorially symmetric and has a poloidal and a toroidal component.
Here, we again see that for $\zeta\lesssim 10^{-3}$, $U_B$ and the rms velocities appear to converge to the same curve at times larger than the value of $t_{\zeta B}$ for each simulation.
This is particularly evident for the toroidal part of the magnetic field, whose magnetic energy is displayed in panel (b).
The rapid decay of the toroidal magnetic energy at early times, on a timescale $t_{\zeta B}$, corresponds to a rearrangement of the magnetic field in which the toroidal component of the magnetic force vanishes (see Sect.~\ref{sec:evolution:GS}).

\subsubsection{Long-term evolution}
\label{sec:results:evolution:longterm}

In Sect.~\ref{sec:results:evolution:FF_convergence}, we have shown that the FF approach is a reliable method for performing simulations for the long-term evolution of the magnetic field under ambipolar diffusion.
Here, we further describe the output of such simulations.

Let us focus on the evolution of magnetic field model II in a star with background density profiles obtained from the HHJ EoS.
This simulation was also analyzed in the previous section, and snapshots of the evolution are given in Fig.~\ref{fig:model II z1e-4 evolution}.
In this figure, we see that the magnetic field evolution occurs in two stages: First, there is a small adjustment to the magnetic field configuration and density perturbations on a timescale $t_{\zeta B}$, in which the FF force, which is initially small due to the chosen value of $\zeta$, becomes even smaller [see Fig.~\ref{fig:model II z1e-4 forces}(a)], indicating that the magnetic field approaches a state of quasi-equilibrium in which Eq.~\eqref{eq:force_balance_1} is satisfied.

\begin{figure}
	\centering
	\includegraphics[width=\linewidth]{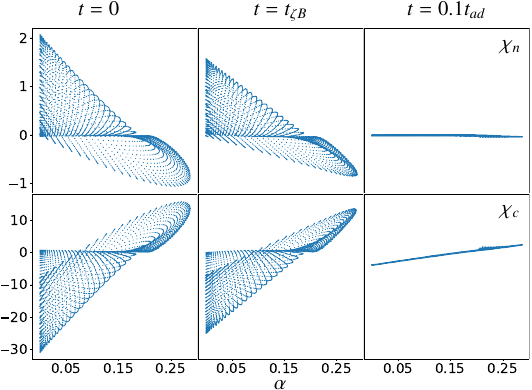}
	\caption{For the simulation of Fig.~\ref{fig:model II z1e-4 evolution}: scatter plot of $\chi_n$ and $\chi_c$ versus $\alpha$ for $t=0$, $t=t_{\zeta B}$, and $t=0.1 t_{ad}$.
	}
	\label{fig:model II z1e-4 versus}
\end{figure}

This quasi-equilibrium state is slowly eroded by ambipolar diffusion. As the charged particles move relative to the neutrons, the magnetic field and particles adjust, moving through a continuum of successive hydromagnetic quasi-equilibria and eventually reaching a final equilibrium state in which $\vec{v}_{n}=\vec{v}_{ad}=0$.
This occurs on a timescale $t_B$, which for the simulation shown in Fig.~\ref{fig:model II z1e-4 evolution} is roughly at $t_B\simeq 0.1 t_{ad}$, as can be appreciated in Fig.~\ref{fig:model II z1e-4 forces}(a).
We observe in Fig.~\ref{fig:z comp model II}(b) and (c) that in this final state, the fluid velocities indeed become very small, thus halting the evolution of the field.

As discussed in Sect.~\ref{sec:evolution:GS}, this kind of axially symmetric equilibrium is called GS equilibrium, in which $\alpha$ is a solution of the GS Eq.~\eqref{eq:GS}, $\chi_c=\chi_c(\alpha)$, and $\chi_n$ is uniform.
This can be observed in the last row of Fig.~\ref{fig:model II z1e-4 evolution} and in Fig.~\ref{fig:model II z1e-4 versus} (at $t=0.1t_{ad}$), suggesting that effectively the magnetic field has evolved towards a GS equilibrium.

\begin{figure}
	\centering
	\includegraphics[width=0.99\linewidth]{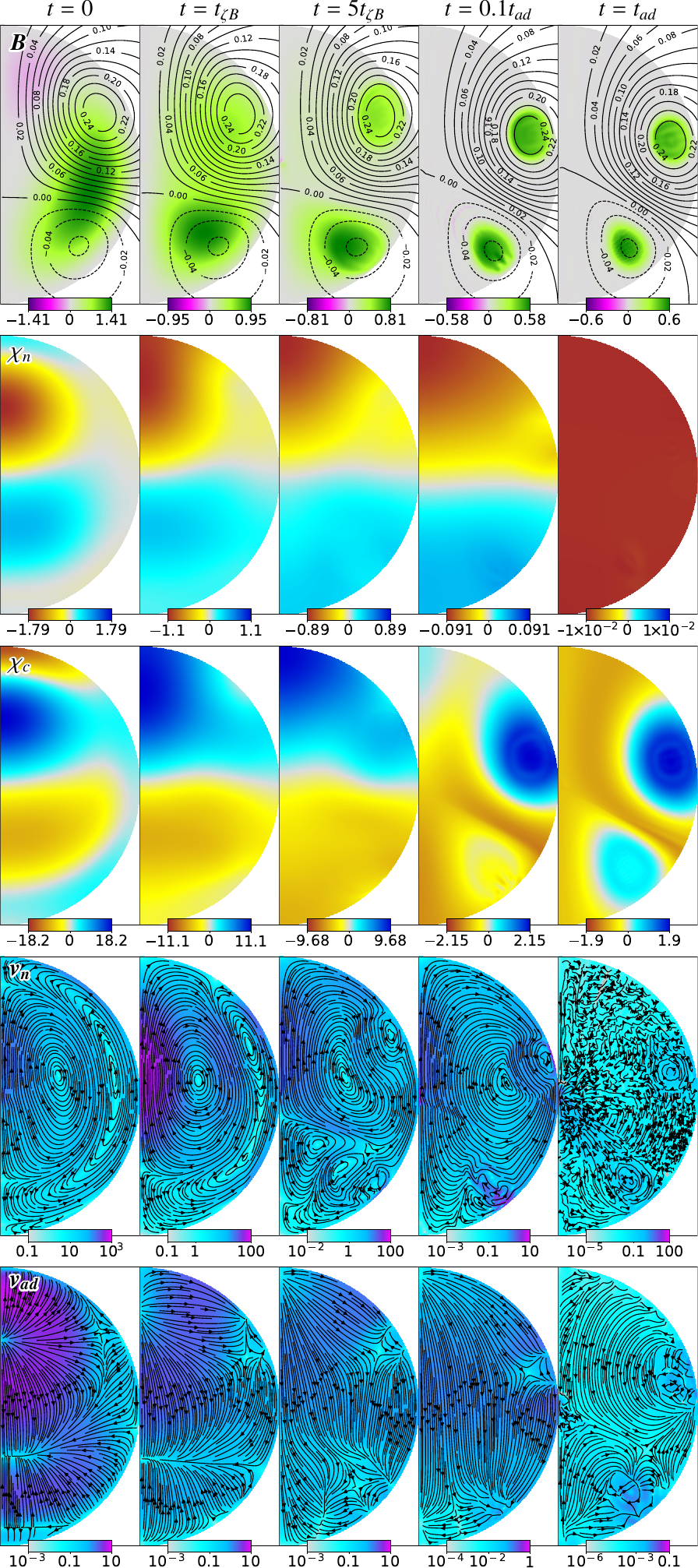}
	\caption{
	Evolution of magnetic field model IV, using $\zeta=10^{-3}$. We used a grid of $N_r=60$ radial steps and $N_\theta=91$ polar steps inside the core and $N_{\text{Exp}}=40$ external multipoles, and the HHJ EoS. From top to bottom: Configuration of the magnetic field, where lines represent the poloidal magnetic field (labeled by the magnitude of $\alpha$) and colors the toroidal potential $\beta$; $\chi_n$ and $\chi_c$, both in units of $\chi_0=B_0^2/4\pi n_0\mu_0$;  poloidal component of the neutron velocity, $\vec{v}_n$, and ambipolar diffusion velocity, $\vec{v}_{ad}$, where arrows represent the direction and colors the magnitude normalized to $R/t_0$. Columns correspond to different times: $t=0$, $t_{\zeta B}$, $5\,t_{\zeta B}$, $0.1\,t_{ad}$, and $t_{ad}$.
	}
	\label{fig:model IV z1e-3 evolution}
\end{figure}

\begin{figure}
	\centering
	\includegraphics[width=\linewidth]{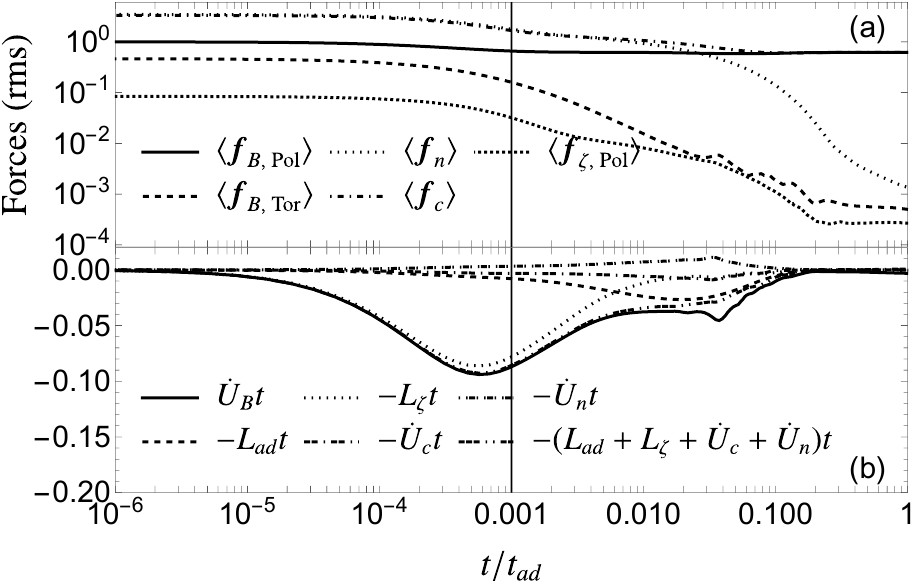}
	\caption{For the simulation of Fig.~\ref{fig:model IV z1e-3 evolution}: (a)~Time evolution of the rms force densities $\langle \vec{f}_{B\text{, Pol}} \rangle$, $\langle \vec{f}_{B\text{, Tor}} \rangle$, $\langle \vec{f}_{n} \rangle$, $\langle \vec{f}_{c} \rangle$, and $\langle \vec{f}_{\zeta\text{, Pol}} \rangle$, all normalized by $\langle \vec{f}_{B\text{, Pol}}(t=0)\rangle$, where $\langle .\rangle$ denotes rms in the volume of the core.
    (b) Energy conservation. The different terms are in units of the total initial magnetic energy.
    Time is in units of $t_{ad}$ for all panels. The vertical line corresponds to $t_{\zeta B}/t_{ad}$.
	}
	\label{fig:model IV z1e-3 forces}
\end{figure}

The conclusions obtained from the time evolution of magnetic field model II remain true for more complex configurations such as model IV, which has poloidal and toroidal components that are both not equatorially symmetric. Snapshots of the time evolution for this model can be observed in Fig.~\ref{fig:model IV z1e-3 evolution}, while the evolution of the rms forces is in Fig.~\ref{fig:model IV z1e-3 forces}(a), and the energy conservation is shown in Fig.~\ref{fig:model IV z1e-3 forces}(b).
We note that the magnetic field is initially not in a state of quasi-equilibrium, as it has a strong toroidal magnetic force.
As expected, this toroidal force becomes progressively weaker, evolving on a timescale $t_{\zeta B}$.
In Fig.~\ref{fig:model IV z1e-3 evolution} at $t=t_{\zeta B}$, we see how quasi-equilibrium is not yet reached as $\beta$ is not a function of $\alpha$. This condition is more clearly satisfied at $t=5t_{\zeta B}$, where $\langle \vec{f}_{B\text{, Tor}} \rangle/\langle \vec{f}_{B\text{, Pol}} \rangle \approx 6\times 10^{-3}$.
Here, we also see that the toroidal magnetic field is confined to regions of closed poloidal field lines, forming ``twisted tori,'' which is a consequence of the condition $\beta=\beta(\alpha)$.
At this point, $\chi_n$ and $\chi_c$ are of the same order. However, as the system evolves due to ambipolar diffusion, $\chi_n$ becomes progressively smaller
and more uniform [see $\langle \vec{f}_{n} \rangle$ in Fig.~\ref{fig:model IV z1e-3 forces}(a)]. As expected, after the GS equilibrium is reached, $\chi_c$ is a function of $\alpha$, and the velocities are much smaller than their initial values, indicating that the system has reached its final equilibrium.

Fig.~\ref{fig:model IV z1e-3 forces}(b) shows energy conservation for this simulation.
There is overall good agreement between the curves of $\dot U_{B}t$, and $-(L_{ad}+L_{\zeta}+\dot U_c+\dot U_n)t$, aside from a very small mismatch around $t\sim 0.05t_{ad}$, which we attribute to numerical viscosity.
Compared to the simulation of Fig.~\ref{fig:model II z1e-4 evolution}, here we see a larger contribution of $L_\zeta$. This is because of two factors.
First, in this simulation, the magnetic field is initially not in quasi-equilibrium.
Thus, there is a noticeable initial rearrangement of the field and the fluids on a timescale $t_{\zeta B}$.
During the initial rearrangement, energy is dissipated due to physical processes that involve the propagation and damping of sound and Alfvén waves, which in our model are mimicked by the FF. 
The rate at which this energy dissipated is accounted for in $L_\zeta$.
In this early stage, we have $v_n\sim f_B/(\zeta n_n)$, thus $L_\zeta\sim \zeta n_n v_n^2R^3 \propto 1/\zeta$.
Therefore, the energy dissipated during the unwinding of the magnetic field lines is $\sim L_\zeta t_{\zeta B}$, which
does not depend on $\zeta$ (as long as it is small enough).
Second, after the star reaches hydromagnetic quasi-equilibrium, $\vec{v}_n$ converges to the real velocity, which does not depend on $\zeta$, and therefore $L_\zeta \propto \zeta$, which is larger in this simulation
($\zeta=10^{-3}$ for the simulation of Fig.~\ref{fig:model IV z1e-3 evolution} and $10^{-4}$ for the simulation of Fig.~\ref{fig:model II z1e-4 evolution}). 

\begin{figure}
	\centering
	\includegraphics[width=\linewidth]{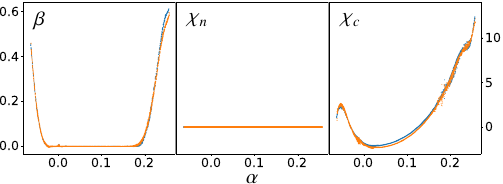}
	\caption{
    Scatter plot of $\beta$, $\chi_n$, and $\chi_c$ versus $\alpha$ for the final equilibrium configuration (at $t=t_{ad}$) of magnetic field model IV for two simulations, using the toy model EoS.
    Blue dots correspond to the results presented in \citet{Castillo2020_TwofluidSimulationsMagnetic}, which are computed using $B_0=1.8\times 10^{17}\text{G}$ (i.~e., $\chi_0=1.67\times 10^{-3}$); 
    orange dots correspond a simulation performed with the current version of the code, using $\zeta=3\times10^{-3}$, $N_r=60$ radial steps, $N_\theta=91$ polar steps inside the core, $N_{\text{Exp}}=27$ external multipoles, and $u=1/2$ (the same parameters as in \citealt{Castillo2020_TwofluidSimulationsMagnetic}).
    All quantities are normalized as described in Table~\ref{tab:normalization}.
	}
	\label{fig:model IV z3e-3 vs}
\end{figure}

The results presented in this section are consistent with those reported in \citet{Castillo2020_TwofluidSimulationsMagnetic}. The time evolution for all simulations presented here qualitatively follows the same stages, and in all cases the final state corresponds to a GS equilibrium.
For completeness, Fig.~\ref{fig:model IV z3e-3 vs} shows a comparison between the final equilibrium state for model IV, computed by the current version of the code, which, as mentioned before, can be rescaled to any realistic value of the magnetic field strength; and the results presented in \citet{Castillo2020_TwofluidSimulationsMagnetic} for the same magnetic field model, in a simulation performed for an unrealistically high magnetic field strength of $B_0=1.8\times 10^{17}\text{G}$.
We see that the final equilibrium configuration reported in \citet{Castillo2020_TwofluidSimulationsMagnetic} is correctly reproduced, and it is found not to be significantly affected by the, now realistic, magnetic field strength used.

\subsubsection{Dependence on the equation of state}
\label{sec:results:evolution:EoS}

In \citet{Castillo2020_TwofluidSimulationsMagnetic}, we performed our simulations using the toy model EoS, described in Sec.~\ref{sec:EoS}.
While it is a simple analytical model, it helped us to correctly analyze the impact of the stable stratification of NS matter on the magnetic field evolution.
However, the question remains if using more realistic density profiles could significantly impact the evolution of the field.
In particular, in real NSs, strong interactions may play a significant role, which could not be observed in the previous simulations using the toy model.

\begin{figure}
\centering
\includegraphics[width=\linewidth]{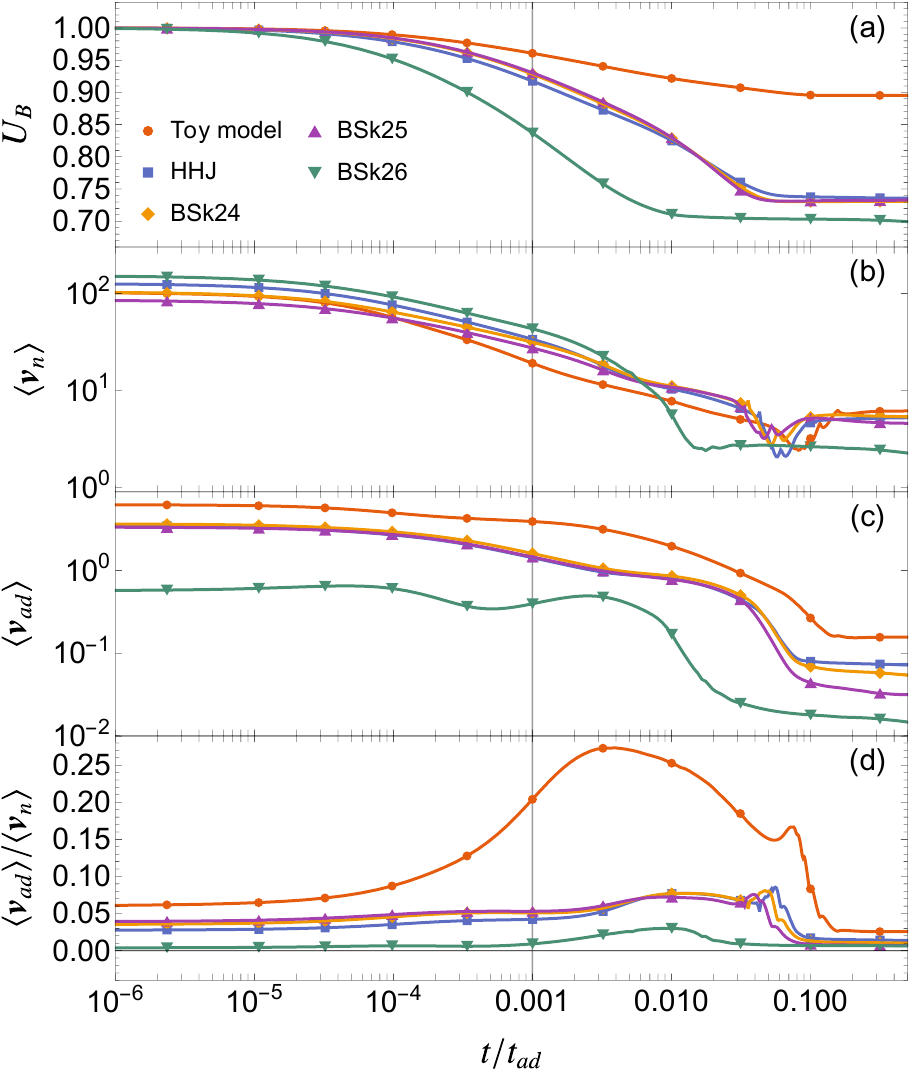}
\caption{Time evolution for the initial purely poloidal magnetic field model III with five different EoSs using $\zeta=10^{-3}$, a grid of $N_r=60$ radial steps, $N_\theta=91$ polar steps inside the core, $N_{\text{Exp}}=40$ external multipoles, and $u=2/3$.
The panels show the following variables: (a)~Magnetic energy in the NS core, normalized by its initial value.
(b) and (c)~rms neutron and ambipolar velocity, $\langle v_{n}\rangle$ and $\langle v_{ad}\rangle$, respectively. (d)~The ratio $\langle v_{ad}\rangle/\langle v_{n}\rangle$.
For each model, the timescales $t_{\zeta B}$ and $t_{ad}$, and the velocities are evaluated considering the stellar parameters of Table~\ref{tab:stellar parameters}.
The vertical line marks the dimensionless value of $\zeta=t_{\zeta B}/t_{ad}$, which is fixed to be the same for all EoSs.
}
\label{fig:EOS comp model III}
\end{figure}

Here, as detailed in Sect.~\ref{sec:EoS}, we use the EoSs HHJ, BSk24, BSk25, and BSk26, all of which consider the effect of strong interactions on the chemical potentials by means of a coefficient $K\neq 0$.
Results for simulations using the purely poloidal magnetic field model III with $\zeta=10^{-3}$ can be seen in Fig.~\ref{fig:EOS comp model III}. 
Here, we see how, for all EoSs considered, the magnetic field decays and eventually reaches an equilibrium state [see panel (a)], but the time at which this occurs for each is not the same.
The toy model reaches equilibrium on a timescale $\sim 0.1\,t_{ad}$, simulations using HHJ, BSk24, and BSk25 reach equilibrium in $\sim 0.05\,t_{ad}$, while for BSk26 equilibrium is reached around $0.01\,t_{ad}$. 
These results are consistent with Eq.~\eqref{eq:t_B}, which states that the timescale for magnetic field evolution under ambipolar diffusion is shorter than $t_{ad}$ by a factor $(\ell_B/\ell_c)^2$. 
In this comparison, only $\ell_c$ varies between EoSs, as seen in Fig.~\ref{fig:n-ratio}(c), while $\ell_B$ remains unchanged.
Note that the toy model yields the smallest value of $\ell_c$ for all $r$, and thus the largest estimate for $t_B$ among the stellar models studied.
HHJ, BSk24, and BSk25 yield similar values of $\ell_c$, while in comparison, BSk26 yields the largest value, thus the shortest estimate for $t_B$.
This is further supported by Fig.~\ref{fig:EOS comp model III}(b)-(d), where we see the rms velocities and their ratio $\langle v_{ad}\rangle/\langle v_{n}\rangle$ as functions of time.
Note that this ratio remains $\sim 0.2$ for most of the evolution when using the radial density profiles from the toy model.
In contrast, for BSk26, this ratio is considerably lower ($\sim 0.02$), consistent with Eq.~\eqref{eq:vn-va}.

\begin{figure}
\centering
\includegraphics[width=\linewidth]{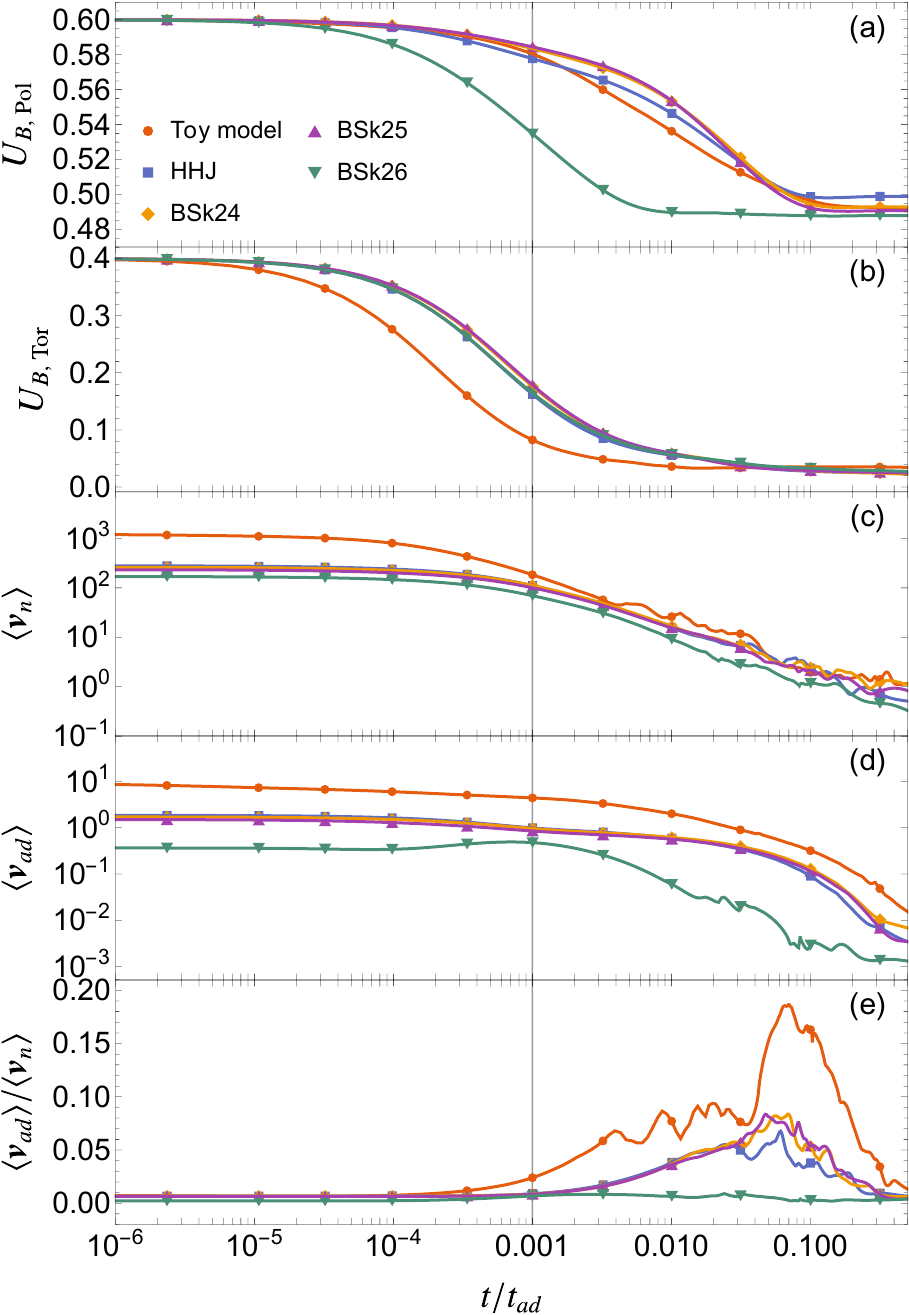}
\caption{
The same as Fig.~\ref{fig:EOS comp model III}, but for model IV with $\zeta=10^{-3}$, $N_r=60$, $N_\theta=91$, $N_{\text{Exp}}=40$, and $u=2/3$.
The variables shown in each panel are: (a) and (b)~Poloidal and toroidal magnetic energy in the NS core, respectively, normalized by the initial value of the total magnetic energy of the NS core.
(c) and (d)~rms neutron and ambipolar velocity, $\langle v_{n}\rangle$ and $\langle v_{ad}\rangle$, respectively. (e)~The ratio $\langle v_{ad}\rangle/\langle v_{n}\rangle$.
}
\label{fig:EOS comp model IV}
\end{figure}

Results for magnetic field model IV, also using $\zeta=10^{-3}$, can be seen in Fig.~\ref{fig:EOS comp model IV}.
Panels (a) and (b) show the evolution of the poloidal and toroidal magnetic energy, respectively; we again see how, for the BSk26 star, the poloidal magnetic energy decays faster than for the other stellar models. 
On the other hand, the toroidal magnetic field decays faster for the toy model than for all the other EoSs, all of which decay on a similar timescale.
The dissipation of the toroidal energy occurs mainly due to the unwinding of the toroidal magnetic field in the regions with poloidal field lines closing outside the core (see Sect.~\ref{sec:evolution:GS}), a process with a typical timescale $t_{\zeta B}$.
This timescale is proportional to $n_n$ (see Eq.~\ref{eq:t_zB}), which decreases towards the surface of the star (note that in the plots, the timescales are evaluated at the center of the star). Thus, the unwinding of the field lines close to the surface is expected to be faster.
Fig.~\ref{fig:n-ratio}(a) shows that close to the crust-core interface, $n_n$ is smaller for the toy model EoS than for the more realistic EoSs, explaining the faster decay of the toroidal field for that case.
Also, the magnetic field in that region is stronger in model IV than in model III, making the difference in $n_n$ have a larger impact on the evolution of the magnetic energy.
Thus, it is not that surprising that the toroidal magnetic energy decays faster for the simulation using the (unrealistic) toy model.
Consequently, the toroidal component of the velocities is expected to be larger for the toy model;
panels (c) and (d) of Fig.~\ref{fig:EOS comp model IV} show that, for the toy model EoS, the rms values of the neutron and ambipolar velocity are larger than for the other EoSs, as these velocity averages are dominated by the toroidal component of the velocities. 
Note that the curves for $\langle v_{ad}\rangle/\langle v_{n}\rangle$ follow a similar trend to the simulation in Fig.~\ref{fig:EOS comp model III}. 
That is, for the toy model EoS, which has the largest ratio $\ell_B/\ell_c$, we have the highest $\langle v_{ad}\rangle/\langle v_{n}\rangle$ ratio. For the more realistic EoSs -- HHJ, BSk24, and BSk25 -- the neutron velocity is considerably larger than the ambipolar velocity, consistent with a ratio $\ell_B/\ell_c \simeq 0.07$. In turn, in the simulation using BSk26, the ratio $\ell_B/\ell_c$ is much smaller, consistent with Fig.~\ref{fig:n-ratio}(c).

\section{Discussion and conclusions}
\label{sec:conclusions}

The evolution of a neutron star’s magnetic field is probably responsible for much of its observed phenomenology. Initially, the magnetic field likely permeates the entire interior of the star. Therefore, understanding its evolution requires studying the key mechanisms driving it, both in the crust and the core. While crustal processes have been extensively studied and are largely understood, the same cannot yet be said for processes in the core.

Assuming a NS core composed only of ``normal'' (non-Cooper-paired) neutrons, protons, and electrons, its magnetic field is carried by a joint motion of the charged particles (protons and electrons), which is similar to that of the neutrons, but crucially not identical, because of the different density profiles entering their continuity equations~(\ref{eq:continuity_n2}) and (\ref{eq:continuity_c2}). Their slight difference (ambipolar diffusion) causes a collisional friction that controls the timescale of this evolution, which is much longer than any of the dynamical timescales of the NS core, such as the periods of sound, gravity, or Alfvén waves. 
Thus, this two-fluid system can be considered to be always near a state of hydromagnetic equilibrium, where the inertial terms in the equations of motion can be ignored. In this regime, the velocity fields are determined from the force-balance and continuity equations for the two fluids (neutrons and charged particles).

Under the assumption of axial symmetry, the velocity fields of neutrons and charged particles were studied semi-analytically by \citet{Gusakov2017_EvolutionMagneticField} and \citet{Ofengeim2018_FastMagneticField} and calculated numerically by \citet{Castillo2020_TwofluidSimulationsMagnetic}, who also used them to simulate the long-term evolution of the magnetic field in the NS core. For this purpose, a small fictitious friction (FF) force on the neutrons was introduced, which simplified the calculation of the velocity fields, allowing to calculate them at every timestep in order to follow the long-term evolution of the magnetic field. 
It was found that, ultimately, this evolution always leads to an equilibrium state in which the magnetic force is balanced by the internal forces of the charged-particle fluid, satisfying the barotropic ``Grad-Shafranov equation,'' while the neutrons reach a diffusive equilibrium that remains unaffected by the magnetic field.

In the present paper, we addressed a few potential weaknesses of the simulations of \citet{Castillo2020_TwofluidSimulationsMagnetic}:

\begin{enumerate}[I.] 
\item The approach used involved two timescales that are very different for realistic magnetic fields. To capture both in the simulations, it was only possible to consider unrealistically high values of the magnetic field ($\gtrsim 10^{17}\text{G}$). 

\item The FF approach used in \citet{Castillo2020_TwofluidSimulationsMagnetic} is conceptually non-trivial. At first glance, it is not evident that it correctly reproduces the velocities needed to perform long-term time-evolving simulations. 

\item The toy model EoS used (see Sect.~\ref{sec:EoS}) correctly captured the qualitative impact of stable stratification on the long-term magnetic evolution. However, the question remained if considering more realistic EoSs would significantly affect the results. 
\end{enumerate}

In order to address item I, we followed \citet{Moraga2024_MagnetothermalEvolutionCores} in neglecting the time derivatives of the density perturbations in the continuity equations, thus eliminating a short timescale that was not interesting for the magnetic field evolution. This left only two fundamental timescales ($t_{\zeta B}$ and $t_{ad}$) that are both $\propto B^{-2}$, allowing to scale the simulation results to any relevant magnetic field strength. 

Item II was addressed analytically by showing that the exact chemical potential perturbations and velocity fields for a given magnetic field configuration correspond to the zeroth-order term in an expansion of the equations with FF in powers of the coefficient $\zeta$. It was further addressed by comparing numerical calculations of the same quantities for both the exact equations (following an explicit and a pseudo-spectral approach) and those with FF, using different values of $\zeta$; and finding good agreement for small $\zeta$.
These calculations were done using our grid-based code developed in-house and compared against its equivalent implementation using \texttt{Dedalus} code.
Moreover, by following the evolution for different values of $\zeta$ we also find convergence for sufficiently small values.
However, we note that the FF approach allows to impose boundary conditions at the crust-core interface, such as no particle flows through the latter, whereas in the equations without FF this is only possible for specific magnetic field configurations satisfying constraints that may not continue to be satisfied throughout the evolution.

From the semi-analytical velocities obtained using the explicit and pseudo-spectral approaches, we provide further evidence supporting the fact that the ambipolar velocity is typically {\it much smaller} than the neutron velocity (in particular, for EoS BSk26, $\langle v_{ad}\rangle/\langle v_n\rangle \sim 0.02$) and that their ratio decreases for more complex magnetic fields.

Finally, item III was addressed by evolving the magnetic field for different state-of-the-art EoSs as well as the toy-model EoS and finding qualitative agreement, with the main quantitative difference being a faster evolution of the poloidal magnetic field for EoSs with smaller composition gradients.

In summary, the present work both validates the previous one by supporting the correctness of the FF approximation and improves on it by neglecting the time derivatives of the density perturbations and including realistic EoSs.

There are, however, several ingredients not considered in this paper 
that should be considered in future work:

\begin{enumerate}
\item Temperature evolution: As the temperature evolves in parallel with the magnetic field, it gradually changes the values of microphysical coefficients such as collisional couplings, thus changing the timescale for the evolution of the magnetic field. On the other hand, magnetic dissipation mechanisms affect the thermal evolution of the neutron star. This magneto-thermal evolution in the NS core is being considered in \citet{Moraga2024_MagnetothermalEvolutionCores} and in Moraga et al., in preparation.

\item MHD instabilities: It is unclear if the GS equilibrium remains stable (or even if it is ever reached) if full three-dimensional motions are allowed, as there are many non-axially symmetric instabilities that may destabilize such equilibria \citep{Tayler1973_AdiabaticStabilityStars,Markey1973_AdiabaticStabilityStars,Wright1973_PinchInstabilitiesMagnetic, Flowers1977_EvolutionPulsarMagnetic}.
Moreover, in this state, the Lorentz force is balanced only by the charged-particle fluid consisting of protons and electrons, which behaves as a barotropic fluid. Numerical simulations suggest that there are no stable equilibria in barotropic stars \citep{Braithwaite2009_AxisymmetricMagneticFields,Akgun2013_StabilityMagneticFields,Lander2012_AreThereAny, Mitchell2015_InstabilityMagneticEquilibria,Becerra2022_EvolutionRandomInitial,Becerra2022_StabilityAxiallySymmetric}
and that even axially symmetric instabilities can be triggered, destabilizing the field \citep{Becerra2022_StabilityAxiallySymmetric}. These instabilities might still be triggered in the two-fluid system at hand, although slowed by the collisional friction between the two fluids.

\item Muons: Adding additional species such as muons makes the charged-particle component non-barotropic, leading to less constrained equilibria that are not GS.
Thus, even if magnetic fields in barotropic stars are always unstable, muons might stabilize them by stably stratifying the charged particle fluid.

\item Superfluidity/superconductivity: At typical core temperatures, NS matter is expected to be both superfluid and superconducting. Numerical simulations have begun incorporating these effects in simplified models (see, e.g., \citealt{Bransgrove2018_MagneticFieldEvolution, Bransgrove2025_GiantHallWaves}); note, however, that some of the microphysical input used in these studies has been criticized \citep{Gusakov2019_ForceProtonVortices}.
Nucleon superfluidity and superconductivity not only influence the kinetic coefficients governing magnetic field dissipation but also lead to the formation of quantized neutron vortices and magnetic flux tubes. These effects can significantly impact the magnetic field evolution described here, altering both the characteristic timescales and the structure of the final equilibrium configuration -- if such a configuration exists (see \citealt{Gusakov2020_MagneticFieldEvolution} for a discussion).

\item The crust: There have been few attempts to follow the coupled evolution of the core and the crust.
One of them is the work of \citet{Vigano2021_MagnetothermalEvolutionNeutron}, in which the magneto-thermal evolution of the crust is coupled to the evolution of the core, including ambipolar diffusion (although neglecting neutron motion).
The recent work of \citet{Skiathas2024_CombinedMagneticField} takes a similar approach.
However, in both cases, the treatment of the core is still simplified.
Thus, more realistic simulations would be desirable.
Whether the magnetic field evolution is faster in the core or crust depends mainly on the field's strength and the crust's compositional impurity and whether the matter in the core has undergone a superfluid/superconducting transition \citep{Cumming2004_MagneticFieldEvolution, Pons2007_MagneticFieldDissipation,Gourgouliatos2016_MagneticFieldEvolution,Akgun2018_CrustMagnetosphereCoupling}.
If the evolution is slower in the crust, it could stabilize the magnetic field in the core, leading to different kinds of equilibria. 
\end{enumerate}

\section*{Acknowledgements}
This work was supported by the CAS-ANID fund N°CAS220008 (F.C.), FONDECYT Projects 11230837 (F.C.), 1201582 (A.R.), and 1240697 (J.A.V.), and ANID doctoral fellowship 21210909 (N.M.). M.G.~acknowledges support from the Ministry of Science and Higher Education of the Russian Federation under the state
assignment FFUG-2024-0002 of the Ioffe Institute. N.M. and F.C. thank the University of Leeds for their hospitality.


\bibliographystyle{aa}
\bibliography{references} % if your bibtex file is called example.bib

\end{document}